\documentclass[reprint, aps, 
superscriptaddress, amsmath, amssymb]{revtex4-2}

\usepackage{graphicx}% Include figure files
\usepackage{bm}% bold math

\usepackage{textcomp,marvosym}

\usepackage{bm}
\usepackage{ulem}
\usepackage{url}
\usepackage{xcolor}
\usepackage{lineno}

\begin{document}

\title{Setting of the Poincar\'e section for accurately calculating the phase of rhythmic spatiotemporal dynamics}

\author{Takahiro Arai}
\email{araitak@jamstec.go.jp}
\affiliation{Center for Mathematical Science and Advanced Technology, Japan Agency for Marine-Earth Science and Technology, Yokohama 236-0001, Japan}

\author{Yoji Kawamura}
\affiliation{Center for Mathematical Science and Advanced Technology, Japan Agency for Marine-Earth Science and Technology, Yokohama 236-0001, Japan}

\author{Toshio Aoyagi}
\affiliation{Graduate School of Informatics, Kyoto University, Yoshida-Honmachi, Sakyo-ku, Kyoto 606-8501, Japan}

\date{\today
}

\clearpage
\begin{abstract}
The synchronization analysis of limit-cycle oscillators is prevalent in many fields, including physics, chemistry, and life sciences. 
It relies on the phase calculation that utilizes measurements.
However, the synchronization of spatiotemporal dynamics cannot be analyzed because a standardized method for calculating the phase has not been established.
The presence of spatial structure complicates the determination of which measurements should be used for accurate phase calculation.
To address this, we explore a method for calculating the phase from the time series of measurements taken at a single spatial grid point.
The phase is calculated to increase linearly between event times when the measurement time series intersects the Poincar\'e section. 
The difference between the calculated phase and the isochron-based phase, resulting from the discrepancy between the isochron and the Poincar\'e section, is evaluated using a linear approximation near the limit-cycle solution.
We found that the difference is small when measurements are taken from regions that dominate the rhythms of the entire spatiotemporal dynamics.
Furthermore, we investigate an alternative method where the Poincar\'e section is applied to the time series obtained through orthogonal decomposition of the entire spatiotemporal dynamics. 
We present two decomposition schemes that utilize the principal component analysis.
For illustration, the phase is calculated from the measurements of spatiotemporal dynamics exhibiting target waves or oscillating spots, simulated by weakly coupled FitzHugh-Nagumo reaction-diffusion models.

\end{abstract}

\maketitle

%%%%%%%%%%%%%%%%%%%%%%%%%%%%%%%%%%%%%%%%%%%%%%%%%%%%%%%%%%%%%
%%%%%%%%%%%%%%%%%%%%%%%%%%%%%%%%%%%%%%%%%%%%%%%%%%%%%%%%%%%%%
%%%%%%%%%%%%%%%%%%%%%%%%%%%%%%%%%%%%%%%%%%%%%%%%%%%%%%%%%%%%%

\section{INTRODUCTION}

\par  
Researchers across various disciplines, including physics, chemistry, and life sciences, have shown significant interest in the dynamics of coupled self-sustained oscillators.
According to the phase reduction theory~\cite{Winfree_1980, Kuramoto_1984}, a multi-dimensional nonlinear system can be simplified to a phase equation with a single phase variable representing the oscillator state. 
This equation describes how the rhythm of oscillations varies due to coupling functions.
This theory underpins an inverse problem framework that reveals causality from measurements using a straightforward description.
Assuming that the system consists of coupled oscillators, this framework allows us to characterize the variations in oscillatory rhythms through phase equations.

\par 
The inverse problem of identifying the direction of coupling~\cite{Rosenblum_2001}, phase sensitivity function~\cite{Galan_2005, Ermentrout_2007, Ota_2009, Imai_2017, Cestnik_2017, Cestnik_2018}, or phase coupling function~\cite{Miyazaki_2006, Tokuda_2007, Kralemann_2007, Kralemann_2008, Stankovski_2012, Duggento_2012, Kralemann_2013, Ota_2014, Stankovski_2017, Matsuki_network_2024} requires developing methods to calculate the phase time series from measurements. 
This is because the inverse problem utilizes the phase time series to obtain phase equations that incorporate the phase coupling function between the oscillators and the phase sensitivity function, which quantifies linear response characteristics of the phase to weak perturbations.
Therefore, accurately calculating the phase is crucial for understanding the properties of systems.
A straightforward method for phase calculation involves linearly interpolating the phase over one period, measured using the Poincar\'e section.
Recent studies have developed methods to calculate the phase more accurately~\cite{Kralemann_2007, Kralemann_2008, Revzen_2008, Cestnik_2017, Cestnik_2018, Rosenblum_2021, Wilshin_2022}. 
Furthermore, many studies employ techniques beyond the phase reduction theory such as the Hilbert transform~\cite{Gengel_2022, Matsuki_2023}, Koopman operator~\cite{Namura_2022, Hashimoto_data-driven_2022, Shirasaka_phase-amplitude_2017, Mauroy_global_2018}, and autoencoder~\cite{Fukami_data-driven_2024, Yawata_phase_2024}. 
These studies allow the phase to capture fluctuations within a single period caused by continuous perturbations, including noise, coupling functions, and external perturbations.
The development of the aforementioned method has provided a valuable tool for uncovering the mechanisms of synchronization in real-world systems~\cite{Kralemann_2013, Stankovski_time-varying_2017, Stankovski_neural_2017, Onojima_dynamical_2018, Suzuki_bayesian_2018, Ota_interaction_2020, Arai_extracting_2022, Arai_interleg_2023, Furukawa_bayesian_2024}.

\par
Many studies have reported that synchronization occurs not only in oscillators but also in spatiotemporal dynamics.
For example, in atmospheric and oceanic circulation~\cite{Stein_2011, Stein_2014}, synchronous phenomena are observed between opposite sides of the globe, such as the sea surface temperatures of the Kuroshio Current and the Gulf Stream~\cite{Kohyama_2021, Cessi_2021}, as well as the atmospheric variability patterns of the Arctic Oscillation and the Antarctic Oscillation~\cite{Tachibana_2018}.
This discovery prompted us to create a method for analyzing the synchronization mechanism underlying spatiotemporal dynamics.
The phase reduction theory has been broadened beyond its original application to limit-cycle oscillator, now encompassing collective oscillations of dynamical elements~\cite{Kori_2009, Kawamura_collective_2011, Kawamura_collective_2017, Nakao_2018} and spatiotemporal dynamics, such as the oscillatory convection~\cite{Kawamura_collective_2013, Kawamura_noise-induced_2014, Kawamura_phase_2015}, reaction-diffusion system~\cite{Nakao_2014, Kawamura_optimizing_2017}, periodic flow~\cite{Taira_phase-response_2018, Iima_jacobian-free_2019, Iima_phase_2021, Kawamura_adjoint-based_2022, Godavarthi_optimal_2023, Iima_optimal_2024}, and beating flagella~\cite{Kawamura_phase_2018}.
As a result, there is a growing expectation for the development of an inverse method to estimate phase equations that characterize these types of dynamics.
However, such an inverse method for spatiotemporal dynamics remains undeveloped due to the lack of established techniques for calculating the phase from measurements.
Due to the spatial structure inherent in spatiotemporal dynamics, several considerations arise for phase calculation:
$(\mathrm{i})$ determining the optimal locations for fixed-point observations of the dynamics;  $(\mathrm{ii})$ exploring the use of modes obtained from spatiotemporal dynamics through decomposition techniques such as principal component analysis (PCA).
Calculating the phase of spatiotemporal dynamics from measurements presents a complex challenge, and this issue has not yet been addressed in existing research.
This study aims to develop a method that allows for the calculation of phase in spatiotemporal dynamics, analogous to the approach used for limit-cycle oscillators.
The method for calculating the phase will be crucial for uncovering causality between spatiotemporal dynamics and will offer a clearer, more intuitive understanding of these systems.

\par 
In this study, we explore two methods for calculating the phase of the spatiotemporal dynamics: one method relies on measurements taken at a single spatial grid point, while the other method utilizes measurements from all spatial grid points.
First, we examine the method for calculating the phase using measurements taken at a single spatial grid point.
This method involves a straightforward technique: measuring the period of spatiotemporal dynamics using the Poincar\'e section applied to the measurement time series, and then linearly interpolating the phase over one period.
The accuracy of phase calculation is influenced by the measurement position.
Thus, we developed an approach to optimize the measurement position for accurate phase calculation.
The difference between the calculated phase and the isochron-based phase can be evaluated using a linear approximation in the vicinity of the limit-cycle solution~\cite{Nakao_2014}.
To illustrate this approach, we provide an example by calculating the phase of spatiotemporal dynamics simulated by coupled FitzHugh-Nagumo (FHN) reaction-diffusion models. 
By selecting suitable parameters, the model can generate various spatiotemporal rhythmic patterns~\cite{Hastings_1976, Hagberg_1994, Nomura_1996, Yanagita_2008}. 
In this study, we simulated target waves and oscillating spots.
Next, we investigate the method for calculating the phase using measurements from all spatial grid points.
In this method, the phase is calculated by linear interpolation, with the Poincar\'e section applied to a one-dimensional time series obtained from the orthogonal decomposition of the spatiotemporal dynamics. 
We propose decomposition schemes using PCA and demonstrate how to calculate the phase of spatiotemporal dynamics through numerical simulation with the FHN reaction-diffusion model.

\par
This present paper is organized as follows: 
In Sec.~\ref{sec:simulation_model}, the FHN reaction-diffusion model is introduced as a simulation model (Sec.~\ref{sec:model_FHNRD}). 
It includes a detailed explanation of two distinct spatiotemporal oscillatory patterns, the target wave (Sec.~\ref{sec:model_TW}) and oscillating spot (Sec.~\ref{sec:model_OS}).
We focus on the method that relies on measurements from a single spatial grid point in Sec.~\ref{sec:1d}. 
This section covers the process of calculating the phase through linear interpolation (Sec.~\ref{sec:1d_phasecal}) and outlines an effective approach for selecting the optimal measurement position (Sec.~\ref{sec:1d_position}). 
Next, we demonstrate the application of our method to spatiotemporal dynamics exhibiting target waves (Sec.~\ref{sec:1d_TW}) and oscillating spots (Sec.~\ref{sec:1d_OS}).
In Sec.~\ref{sec:PCA}, we shift focus to methods involving measurements from all spatial grid points.
This section details orthogonal decomposition (Sec.~\ref{sec:PCA_decomposition}), the phase calculation through linear interpolation (Sec.~\ref{sec:PCA_phasecal}), and the difference between the calculated phase and the isochron-based phase (Sec.~\ref{sec:PCA_correction_term}). 
Then, we propose two orthogonal decomposition schemes utilizing PCA and demonstrate their application in calculating the phase for spatiotemporal dynamics with target waves and oscillating spots (Secs.~\ref{sec:PCA_uv} and ~\ref{sec:PCA_QuQv}).
Finally, Sec.~\ref{sec:discussion} summarizes our results and discusses future research directions related to this study.

%%%%%%%%%%%%%%%%%%%%%%%%%%%%%%%%%%%%%%%%%%%%%%%%%%%%%%%%%%%%%
%%%%%%%%%%%%%%%%%%%%%%%%%%%%%%%%%%%%%%%%%%%%%%%%%%%%%%%%%%%%%
%%%%%%%%%%%%%%%%%%%%%%%%%%%%%%%%%%%%%%%%%%%%%%%%%%%%%%%%%%%%%

\section{SIMULATION MODEL}
\label{sec:simulation_model}
In this section, we explain the FitzHugh-Nagumo (FHN) reaction-diffusion model used for the numerical simulation in this study.
We begin with a concise overview of the reaction-diffusion model  (Sec.~\ref{sec:model_FHNRD}).
Following this, we detail the model setup for both a target-wave solution (Sec.~\ref{sec:model_TW}) and an oscillating spot solution (Sec.~\ref{sec:model_OS}).
The conventional concept of the isochron, which assigns a phase to a subset of the state space in an ordinary differential equation (ODE), is described in Appendix~\ref{appendix:isochron_ODE}.
See Appendix~\ref{appendix:isochron} for the concept of the isochron as applied to a partial differential equation (PDE).

\subsection{A pair of weakly coupled FHN reaction-diffusion models}
\label{sec:model_FHNRD}

\par
We consider a pair of weakly coupled reaction-diffusion models. 
The general form of this dynamical system is described by the following equation:
\begin{align}
    \begin{split}
    \frac{\partial}{\partial t} \bm{X}_1(\bm{r},t)
    = \bm{F}_1(\bm{X}_1, \bm{r}) + D_1 \nabla^2 \bm{X}_1 + \bm{G} (\bm{X}_1, \bm{X}_2),\\
    \frac{\partial}{\partial t} \bm{X}_2(\bm{r},t)
    = \bm{F}_2(\bm{X}_2, \bm{r}) + D_2 \nabla^2 \bm{X}_2 + \bm{G} (\bm{X}_2, \bm{X}_1),
    \end{split}
\label{eq:RDmodel}
\end{align}
where $\bm{X}_i(\bm{r},t) \in \mathbb{R}^N$ represents the state variable of system $i$ at point $\bm{r}$ and time $t$, 
$\bm{F}_i(\bm{X}, \bm{r})$ represents the local reaction dynamics at $\bm{r}$, 
$D_i \nabla^2 \bm{X}_i$ represents the diffusion of $\bm{X}_i$ over the field with a diffusion matrix $D_i$, 
and $\bm{G}(\bm{X}_i, \bm{X}_j)=K[\bm{X}_j(\bm{r},t)-\bm{X}_i(\bm{r},t)]$ represents local and linear mutual couplings with a diagonal matrix $K$ representing the intensity of the mutual coupling. 
We assume that the reaction-diffusion model, when uncoupled ($\bm{G}=\bm{0}$), exhibits a limit-cycle solution with a period $T_i$. 
Additionally, we consider that the mutual coupling is sufficiently weak, such that the state $\bm{X}_i$ remains close to the limit-cycle solution.
As in an ODE case~\cite{Winfree_1980, Kuramoto_1984}, the phase is defined over the basin of attraction of the limit-cycle solution using the concept of isochrons ~\cite{Nakao_2014} (see Appendix~\ref{appendix:isochron_ODE} for the ODE case and Appendix~\ref{appendix:isochron} for the PDE case).
We first assign the phase $\phi_i(t)=\omega_i t$ to the state on the limit cycle, ensuring that the phase increases linearly with a constant frequency $\omega_i := 2\pi/T_i$.
The state on the limit-cycle solution corresponding to $\phi_i$ is represented by $\bm{\chi}_i(\bm{r}, \phi_i)$. 
This state satisfies $\bm{\chi}_i(\bm{r}, \phi_i)=\bm{\chi}_i(\bm{r}, \phi_i+2\pi)$ due to the $2\pi$-periodicity.
Next, we extend the phase assignment to the entire basin of attraction of the limit-cycle solution, enabling us to assign a phase to the state variable $\bm{X}_i$ even when it is not on the limit cycle. 
As a result, the phase $\phi_i$ approximately represents the state $\bm{X}_i$.  
The fluctuation of the phase reflects the changes in the rhythm of spatiotemporal dynamics due to the mutual coupling $\bm{G}$.
Our goal is to establish a method to calculate the phase time series from measurements of the spatiotemporal dynamics without requiring knowledge of the governing equation.

\par 
We consider a one-dimensional FHN reaction-diffusion model. 
The variable $\bm{r}$ in Eq.~(\ref{eq:RDmodel}) is replaced with $x$, representing a point in one-dimensional space.
The state variable, diffusion coefficient, local reaction dynamics, and the mutual coupling in Eq.~(\ref{eq:RDmodel}) are defined as follows:
\begin{align}
    \label{eq:FHNRDmodel}
    \begin{split}        
    &\bm{X}_i(x,t) = 
    \left(  \begin{array}{c}
    u_i \\
    v_i 
    \end{array} \right), \\
    &\bm{F}_i(\bm{X}_i,x) =
    \left(  \begin{array}{c}
    u_i(u_i-\alpha)(1-u_i)-v_i\\
    \tau_i^{-1} (u_i-\gamma v_i)    
    \end{array} \right),  \\
    &D_i = 
    \left( 
    \begin{array}{cc}
        \kappa_i & 0\\
        0 & \delta_i
    \end{array}
    \right), 
    \hspace{5mm}
    \bm{G}(\bm{X}_i, \bm{X}_j) = 
    K  \left( 
    \begin{array}{c}
        u_j - u_i\\
        v_j - v_i
    \end{array} \right),
    \end{split}
\end{align}
where $u_i=u_i(x,t)$ and $v_i=v_i(x,t)$ are activator and inhibitor variables, respectively.
We denote the limit-cycle solution of the model without mutual coupling ($\bm{G}=\bm{0}$) as $\bm{\chi}_i(x, \phi_i)=\left( \chi_{u_i}(x, \phi_i), \chi_{v_i}(x, \phi_i) \right)$.
The spatiotemporal dynamics can exhibit various typical patterns, such as circulating pulses on a ring, oscillating spots, target waves, and rotating spirals by setting the parameters $\alpha$, $\tau_i$, $\gamma_i$, and the diffusion coefficient $\kappa_i$ and $\delta_i$ appropriately~\cite{Hastings_1976, Hagberg_1994, Nomura_1996, Yanagita_2008}.
The parameter $\alpha = \alpha(x)$ is space dependent. 
In this study, we simulate rhythmic patterns of target waves and oscillating spots.

\subsection{Target-wave solution}
\label{sec:model_TW}

\par
As the first example of typical spatiotemporal dynamics, we consider target waves.
To create a pacemaker region, the parameter $\alpha$ is assumed to possess heterogeneity, i.e., $\alpha (x)= \alpha_0 + (\alpha_1-\alpha_0) \exp (-r^4/r_0^4)$, where $r=|x-x_0|$ represents the distance from the center of the pacemaker region, and $r_0$ is the pacemaker region’s radius.
Specifically, $\alpha(x) \to \alpha_1$ as $r \to 0$, and $\alpha(x) \to \alpha_0$ as $r \to \infty$.
The parameters defining the pacemaker region are $r_0=10$, $x_0=80$, $a_0=0.1$, and $a_1=-0.1$.
Other parameters are $\gamma=2.5$, $\tau_1^{-1}=0.005$, $\tau_2^{-1}=0.0055$, $\kappa_1=\kappa_2=0.15$, and $\delta_1=\delta_2=0$. 
The periods of the limit-cycle solutions are $T_1 \simeq 204.6$ and $T_2 \simeq 189.4$.
The coupling intensity is $K=\mathrm{diag}(5.0 \times 10^{-4}, 0)$.
For the numerical simulations, we use a one-dimensional system of size $L=100$ with no-flux boundary conditions, discretized into spatial grids with $\Delta x=L/2^8$. 
Time integration is performed using the explicit Heun method with a time step $\Delta t = 0.01$.

\par
Figure~\ref{fig:fig1}(a) shows the spatiotemporal dynamics of $u_1$ and $v_1$. 
The pacemaker region with radius $r_0$ and center $x=x_0$ is self-oscillatory and rhythmically emits target waves. 
The waves propagate from the pacemaker region outward through the excitable surrounding area.
The limit-cycle solution is depicted in Fig.~\ref{fig:figA1}(a) in Appendix~\ref{appendix:limitcycle}, and Fig.~\ref{fig:figA1}(b) in Appendix~\ref{appendix:limitcycle} shows the phase sensitivity function, $\bm{Q}_i(x, \phi_i)= (Q_{u_i}(x,\phi_i), Q_{v_i}(x,\phi_i))$, which quantifies linear response characteristics of the phase to weak perturbation.
The phase sensitivity function is localized at the pacemaker region (near $x=80$), indicating that this region primarily influences the rhythm of the entire system.

\subsection{Oscillating spot solution}
\label{sec:model_OS}

\par
The second example of typical spatiotemporal dynamics involves oscillating spots.
The parameter $\alpha$ varies spatially and is defined as $\alpha (x)= \alpha_0 + (\alpha_1-\alpha_0) (2x/L-1)^2$, making $\alpha$ largest at the center ($x=L/2$) and smallest at the boundaries ($x=0,L$).
Other parameters are $a_0=-1.1$, $a_1=1.6$, $\gamma=2.0$, $\tau_1^{-1}=0.03$, $\tau_2^{-1}=0.028$, $\kappa_1=1.0$, $\kappa_2=0.9$, $\delta_1=2.5$, and $\delta_2=2.4$. 
The periods of the limit-cycle solutions are $T_1 \simeq 195.3$ and $T_2 \simeq 212.2$.
The coupling intensity is $K=\mathrm{diag}(1.0 \times 10^{-4}, 0)$.
In the numerical simulations, the system is one-dimensional with a size of $L=80$ with no-flux boundary conditions. 
It is discretized into spatial grids with $\Delta x=L/2^8$.
Time integration is performed using the explicit Heun method with a time step $\Delta t = 0.01$.

\par
Under these conditions, oscillating spots constrained at the center are generated.
Figure~\ref{fig:fig1}(b) shows the spatiotemporal dynamics of $u_1$ and $v_1$. 
The oscillating spot, characterized by large values of $u_i$ and $v_i$, can be seen rhythmically expanding and contracting.
The fronts of the oscillating spot for $u_i$ appear sharply.
Therefore, the time series of $u_i(x_p,t)$ measured at $x_p$ within the region where the fronts oscillate, shows intermittent sharp increases and decreases.
In contrast, in other regions, the time series of $u_i(x_p,t)$ exhibits only slight variations without abrupt changes.
The limit-cycle solution and the phase sensitivity function are shown in Figs.~\ref{fig:figA1}(c) and~\ref{fig:figA1}(d) in~Appendix~\ref{appendix:limitcycle}, respectively.
The phase sensitivity function is localized at the fronts of the spot. 
This fact indicates that the region where the front exists influences the rhythm of the entire system.

%%%%%%%%%%%%%%%%%%%%%%%%%%%%%%%%%%%%%%%%%%%%%%%%%%%%%%%%%%%%%
%%%%%%%%%%%%%%%%%%%%%%%%%%%%%%%%%%%%%%%%%%%%%%%%%%%%%%%%%%%%%
%%%%%%%%%%%%%%%%%%%%%%%%%%%%%%%%%%%%%%%%%%%%%%%%%%%%%%%%%%%%%

\section{CALCULATING THE PHASE FROM A MEASUREMENT ON A SINGLE SPATIAL GRID POINT}
\label{sec:1d}

\par
This section details the method for calculating the phase using a Poincar\'e section applied to the time series of measurements of spatiotemporal dynamics taken at a fixed position.
We first describe the linear interpolation technique for phase calculation (Sec.~\ref{sec:1d_phasecal}) and the approach for selecting the optimal measurement position to ensure accurate phase calculation (Sec.~\ref{sec:1d_position}).
Subsequently, we demonstrate the phase calculation process through numerical simulations of the FHN reaction-diffusion model, which exhibits target waves (Sec.~\ref{sec:1d_TW}) and oscillating spots (Sec.~\ref{sec:1d_OS}).

\subsection{Phase calculation by linear interpolation}
\label{sec:1d_phasecal}

\par
We calculate the phase by applying a Poincar\'e section to the time series of $u_i(x_p,t)$, which is measured at a specific spatial grid point of $x=x_p$.
Let us consider an example of spatiotemporal dynamics exhibiting target waves.
Figure~\ref{fig:fig2}(a) shows the time series of $u_i(x_p, t)$ measured at $x_p \simeq 15, 30, 80$.
As shown in this figure, the Poincar\'e sections are applied to each time series, recording the time $t_{i,j}^{x_p}$ when $u_i(x_p, t)$ intersects the Poincar\'e section (from negative to positive) for the $j$th time.
The phase $\theta_i^{x_p}$ is then calculated from the set of times, $\{t_{i,j}^{x_p}\}_j$, ensuring that it satisfies $\theta_{i,j}^{x_p}(t_{i,j}^{x_p})=2\pi j$.
To make the time series of $\theta_i^{x_p}(t)$, we employ linear interpolation of the phase as follows:
\begin{align}
    \label{eq:1d_theta}
    \theta_i^{x_p}(t) = 2\pi j + 2\pi \frac{t-t_{i,j}^{x_p}}{t_{i,j+1}^{x_p}-t_{i,j}^{x_p}} \hspace{5mm} 
    (t_{i,j}^{x_p} \leq t \leq t_{i,j+1}^{x_p}).
\end{align}
Figure~\ref{fig:fig2}(b) presents the time series of $\theta_i^{x_p}$ calculated from the time series of $u_i(x_p, t)$ for $x_p \simeq 30$.
The interval of the time grid for the interpolation is $0.1$.
Figure~\ref{fig:fig2}(c) shows histograms of $\theta_1^{x_p}(t)-\theta_2^{x_p}(t)$ for $x_p \simeq 15, 30, 80$.
The figure indicates the dependency of $\theta_i^{x_p}$ on $x_p$.
For example, the histograms change as $x_p$ varies.

\subsection{Approach to determine the position to measure}
\label{sec:1d_position}

\par
We also calculate the isochron-based phase $\phi_i(t)$.
Generally, there is a difference between the two phases, $\theta_i^{x_p}$ and $\phi_i$, which can be evaluated using a linear approximation near the limit-cycle solution.
The method for calculating the phase using the linear approximation in the ODE case is described in Appendix~\ref{appendix:isochron_ODE}, while the corresponding approach for the PDE case is described in Appendix~\ref{appendix:isochron}.
Here, let us consider the general reaction-diffusion model described in Eq.~(\ref{eq:RDmodel}).
We assume that a Poincar\'e section is applied to the time series of $X_{i,n_p}(x_p, t)$, which is the $n_p$th entry of $\bm{X}_i(x_p, t)$. We record the time $t_{i,j}^{x_p}$ when $X_{i,n_p}(x_p,t)$ intersects the Poincar\'e section for the $j$th time. 
The phase time series is then calculated using Eq.~(\ref{eq:1d_theta}).
Using the phase sensitivity function $\bm{Q}_{i}(x,\phi_i) \in \mathbb{R}^N$, which quantifies linear response characteristics of the phase to weak perturbations, $\phi_i(t_{i,j}^{x_p})$ is calculated from $\theta_i^{x_p}(t_{i,j}^{x_p})$ as follows:
\begin{align}
    \label{eq:1d_correction_term}
    \begin{split}
    & \phi_i (t_{i,j}^{x_p}) 
    = \theta_i^{x_p}(t_{i,j}^{x_p}) +  c_{i,j}^{x_p} 
    = 2 \pi j +  c_{i,j}^{x_p}, \\
    & c_{i,j}^{x_p} = \sum_{\substack{(m,n) \\ \neq (m_p, n_p)}}^{(M,N)} 
    Q_{i,n}(x_m, 0) \Delta_{i,n}(x_m, t_{i,j}^{x_p})
    \Delta x_m, \\
    & \Delta_{i,n}(x,t_{i,j}^{x_p}) := X_{i,n}(x,t_{i,j}^{x_p})-\chi_{i,n}(x,0),
    \end{split} \\
    & \Delta x_m := 
    \begin{cases}
        \Delta x / 2 & (m=0,M), \\
        \Delta x     & (\mathrm{otherwise}), \\
    \end{cases} \nonumber
\end{align}
where $X_{i,n}(x_m,t_{i,j}^{x_p})$, $\chi_{i,n}(x_m, 0)$, and $Q_{i,n}(x_m, 0)$ are the $n$th entry of $\bm{X}_{i}(x_m,t_{i,j}^{x_p})$, $\bm{\chi}_{i}(x_m, 0)$, and $\bm{Q}_{i}(x_m, 0)$, respectively, and $\Delta_{i,n}$ is the deviation of $X_{i,n}$ from the state of $\phi_i=0$ on the limit-cycle solution.
The integration about $x$ is expressed by discrete representation with the spatial grid $x_m= m \Delta x$~$(m=0,1,\ldots,M)$ that satisfies $x_{m_p} = x_p$ and $x_M=L$.
The entry of $(m_p, n_p)$ are removed from the summation in Eq.~(\ref{eq:1d_correction_term}) since the two values, $X_{i,n_p}(x_p, t_{i,j}^{x_p})$ and $\chi_{i,n_p}(x_p, 0)$, have the same value on the Poincar\'e section, i.e., $\Delta_{i,n_p}(x_{m_p}, t_{i,j}^{x_p}) = 0$.
We call the variable $c_{i,j}^{x_p}$, which represents the difference between $\theta_i^{x_p}(t_{i,j}^{x_p})$ and $\phi_i(t_{i,j}^{x_p})$, the correction term.
Figure~\ref{fig:fig3}(a) explains how the correction term occurs.
The figure illustrates that the two states, $\bm{\chi}_i(x, 0)$ and $\bm{X}_i(x,t_{i,j}^{x_p})$, lie on the same Poincar\'e section $P$ but on different isochrons, $I(0)$ and $I(c_{i,j}^{x_p})$. 
Generally, these states are located on distinct isochrons unless the Poincar\'e section is the same as $I(0)$. 
The correction term accounts for the difference between the phases associated with the two isochrons.

\par
We rewrite Eq.~(\ref{eq:1d_correction_term}) for the FHN reaction-diffusion model as follows:
\begin{align}
    \label{eq:1d_correction_term2}
    \begin{split}
    & \phi_i (t_{i,j}^{x_p}) = 2 \pi j +  c_{i,j}^{x_p}, \\
    & c_{i,j}^{x_p}  = \sum_{\substack{m \neq {m_p}}}^M Q_{u_i}(x_m,0) \Delta_{u_i}(x_m, t_{i,j}^{x_p}) \Delta x_m \\
    & \hspace{3em} + \sum_{m=0}^M Q_{v_i}(x_m,0) \Delta_{v_i}(x_m, t_{i,j}^{x_p}) \Delta x_m,
    \end{split} \\
    & \Delta_{u_i}(x, t_{i,j}^{x_p}) :=  u_i(x, t_{i,j}^{x_p}) - \chi_{u_i}(x, 0), \nonumber \\
    & \Delta_{v_i}(x, t_{i,j}^{x_p}) :=  v_i(x, t_{i,j}^{x_p}) - \chi_{v_i}(x, 0), \nonumber \\
    & \Delta x_m := 
    \begin{cases}
        \Delta x / 2 & (m=0,M), \\
        \Delta x     & (\mathrm{otherwise}), \\
    \end{cases} \nonumber
\end{align}
where $Q_{u_i}(x, \phi_i)$ and $Q_{v_i}(x, \phi_i)$ are the phase sensitivity functions of $u_i$ and $v_i$, respectively, and $\Delta_{u_i}$ and $\Delta_{v_i}$ are the deviations of $u_i$ and $v_i$ from the state of $\phi_i=0$ on the limit-cycle solution, respectively.
The subscript $n$ used in Eq.~(\ref{eq:1d_correction_term}) is removed since the entry of two-dimensional state ($n=1,2$) is represented by variable $u_i$ and $v_i$. 
Additionally, the subscript $n_p$ is removed since the Poincar\'e section applied solely to the time series of $u_i(x_p, t)$.
We then proceed to linearly interpolate the phase $\phi_i(t)$ as follows:
\begin{align}
    \label{eq:1d_phi}
    \phi_i(t) 
        &= \phi_i(t_{i,j}^{x_p}) + (\phi_i(t_{i,j+1}^{x_p})-\phi_i(t_{i,j}^{x_p})) \frac{t-t_{i,j}^{x_p}}{t_{i,j+1}^{x_p}-t_{i,j}^{x_p}}  \nonumber \\
        &= 2 \pi j + c_{i,j}^{x_p}+ (2\pi + c_{i,j+1}^{x_p}-c_{i,j}^{x_p}) \frac{t-t_{i,j}^{x_p}}{t_{i,j+1}^{x_p}-t_{i,j}^{x_p}} \\
        & (t_{i,j}^{x_p} \leq t \leq t_{i,j+1}^{x_p}). \nonumber
\end{align} 
The transformation to the second row is achieved by substituting Eq.~(\ref{eq:1d_correction_term2}) into the first row.
Similar to the interpolation process for $\theta_i^{x_p}(t)$ described in Eq.~(\ref{eq:1d_theta}), the time grid interval for this interpolation is set to $0.1$.

\par 
The correction terms are influenced by the two factors: 
($\mathrm{i}$) deviation from the limit-cycle solution;
($\mathrm{ii}$) heterogeneity of the amplitude of the phase sensitivity function. 
Let us consider calculating $\theta_i^{x_p}$ for spatiotemporal dynamics with target waves using the time series of $u_i(x_p, t)$ under the condition of $x_p \simeq 30$, for example.
Figure~\ref{fig:fig3}(b) shows the distribution of $\Delta_{u_i}(x, t_{i,j}^{x_p})$ and $\Delta_{v_i}(x, t_{i,j}^{x_p})$ as well as $Q_{u_i}(x, 0)$ and $Q_{v_i}(x, 0)$.
The figure indicates that the region with large $|\Delta_{u_i}(x, t_{i,j}^{x_p})|$ and $|\Delta_{v_i}(x, t_{i,j}^{x_p})|$ overlaps the region with large $|Q_{u_i}(x, 0)|$ and $|Q_{v_i}(x, 0)|$.
In such a case, the magnitude of $c_{i,j}^{x_p}$ is supposed to be large according to Eq.~(\ref{eq:1d_correction_term2}).
This overlap can be avoided by setting $x_p$ appropriately since the region with large $|\Delta_{u_i}(x, t_{i,j}^{x_p})|$ and $|\Delta_{v_i}(x, t_{i,j}^{x_p})|$ varies with $x_p$. 
Specifically, $|\Delta_{u_i}(x, t_{i,j}^{x_p})|$ and $|\Delta_{v_i}(x, t_{i,j}^{x_p})|$ near $x_p$ are almost zero.
Therefore, a strategy to reduce the magnitude of $c_{i,j}^{x_p}$ is to choose $x_p$ within the region where the amplitudes of the phase sensitivity functions are large. 
It is known that the phase sensitivity function is localized at the pacemaker region of target waves and the fronts of the oscillating spots~\cite{Nakao_2014}.
In Secs.~\ref{sec:1d_TW} and~\ref{sec:1d_OS}, we illustrate this approach in the FHN reaction-diffusion model, demonstrating it for both target waves and oscillating spots.

\subsection{Case of target waves}
\label{sec:1d_TW}

\par
Figure~\ref{fig:fig1}(a) shows the spatiotemporal dynamics exhibiting target waves.
We calculated the time series of $\theta_i^{x_p}(t)$ from that of $u_i(x_p, t)$ for $x_p \simeq 15, 30, 80$ (the exact values are $x_p= 38 \Delta x, 77 \Delta x, 205 \Delta x$). 
Only $x_p \simeq 80$ belongs to the pacemaker region.
Examples of the time series of $u_i(x_p, t)$ and the Poincar\'e section for each $x_p$ are shown in Fig.~\ref{fig:fig2}(a).
Figure~\ref{fig:fig2}(c) shows that the histograms of $\theta_1^{x_p} - \theta_2^{x_p}$ vary with $x_p$. 
The difference between the histogram of $\theta_1^{x_p} - \theta_2^{x_p}$ and the distribution of $\phi_1 - \phi_2$, calculated from the phase equations (Appendix~\ref{appendix:phase_equation}), is minimal for $x_p \simeq 80$.
Therefore, the approach described in Sec.~\ref{sec:1d_position} appears to be effective.

\par
To verify the effectiveness of the approach, we refer to Fig.~\ref{fig:fig4}(a), which shows the distribution of $\Delta_{u_i}(x, t_{i,j}^{x_p})$ and $\Delta_{v_i}(x, t_{i,j}^{x_p})$, along with $Q_{u_i}(x, 0)$ and $Q_{v_i}(x, 0)$.
The region with large $|\Delta_{u_i}(x,t_{i,j}^{x_p})|$ does not overlap the region with large $|Q_{u_i}(x,0)|$ for $x_p \simeq 80$ although the two regions overlap for $x_p \simeq 15, 30$ (the same is true for $Q_{v_i}$ and $\Delta_{v_i}$). 
Given this result, the magnitude of the correction term is relatively small for $x_p \simeq 80$ and larger for the other $x_p$. 
The statistics of the correction term indicate that the correction term is close to zero only for $x_p \simeq 80$ (Fig.~\ref{fig:fig4}(b)).

\par
Ignoring the correction term impacts the phase calculation results, with the extent of the impact varying according to the magnitude of the correction term.
Figure~\ref{fig:fig4}(c) shows the histograms of $\phi_1 - \phi_2$ calculated on the basis of Eqs.~(\ref{eq:1d_correction_term2}) and (\ref{eq:1d_phi}).
The histograms of $\phi_1-\phi_2$ for $x_p \simeq 15, 30, 80$ are almost similar, although the histogram of $\theta_1^{x_p}-\theta_2^{x_p}$ shown in Fig.~\ref{fig:fig2}(c) varies with $x_p$.
It is evident that the histograms of $\theta_1^{x_p}-\theta_2^{x_p}$ and $\phi_1-\phi_2$ differ significantly for $x_p \simeq 15, 30$ since the magnitude of the correction term is large, whereas they are similar for $x_p \simeq 80$ since the correction term is nearly zero.
Ignoring the correction term degrades the accuracy of phase equation estimation when $x_p$ is not appropriate.
Figures~\ref{fig:figA2}(a) and~\ref{fig:figA2}(b) in Appendix~\ref{appendix:estimation} show the phase equations estimated from the time series of $\theta_i^{x_p}(t)$ and $\phi_i(t)$, respectively.
The phase equations estimated from $\theta_i^{x_p}$ deviate significantly from the true forms for $x_p \simeq 15, 30$, but closely match the true forms for $x_p \simeq 80$. 
In contrast, the phase equations estimated from $\phi_i$ are consistent with the true forms regardless of $x_p$.

\subsection{Case of oscillating spots}
\label{sec:1d_OS}

\par
\begin{figure*}[h]
    \begin{center}
    \includegraphics[scale=1.0]{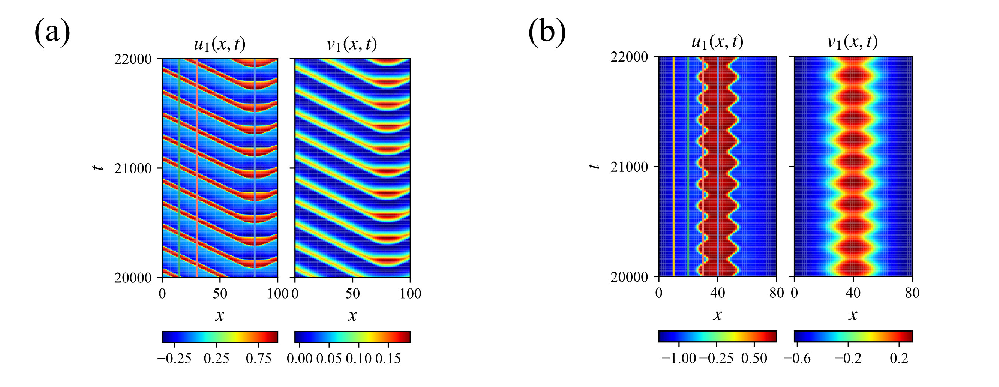}
    \caption{
    The spatiotemporal dynamics simulated by the coupled FHN reaction-diffusion models.
    (a)
    Spatiotemporal dynamics exhibiting target waves.
    For reference, the green, red, and blue vertical lines mark the measurement positions, $x_p \simeq 15, 30, 80$, respectively (see Sec.~\ref{sec:1d_TW}). 
    (b) 
    Spatiotemporal dynamics exhibiting oscillating spots.
    For reference, the yellow, green, red, and blue vertical lines mark the measurement positions, $x_p=10,20,30,40$, respectively (see Sec.~\ref{sec:1d_OS}).
    }
    \label{fig:fig1}
    \end{center}
\end{figure*}

 
We also calculated the phase for the spatiotemporal dynamics of oscillating spots (Fig.~\ref{fig:fig1}(b)).
The time series of $\theta_i^{x_p}(t)$ was calculated from that of $u_i(x_p, t)$ for $x_p=10, 20, 30, 40$. 
Only $x_p=30$ is located in the region containing the spot's front.
Examples of the time series of $u_i(x_p,t)$ and the Poincar\'e sections for each $x_p$ are shown in Fig.~\ref{fig:fig5}(a).
According to Fig.~\ref{fig:fig5}(b), the difference between the histogram of $\theta_1^{x_p}-\theta_2^{x_p}$ and the distribution of $\phi_1-\phi_2$ calculated from the phase equation is the smallest for $x_p=30$. 
Therefore, the approach described in Sec.~\ref{sec:1d_position} is effective for oscillating spots.

\par
Figure~\ref{fig:fig6}(a) shows the distribution of $\Delta_{u_i}(x, t_{i,j}^{x_p})$ and $\Delta_{v_i}(x, t_{i,j}^{x_p})$ as well as $Q_{u_i}(x, 0)$ and $Q_{v_i}(x, 0)$. 
The figure indicates that the region with large $|\Delta_{u_i}(x,t_{i,j}^{x_p})|$ do not overlap the region with large $|Q_{u_i}(x, 0)|$ only for $x_p=30$. (the same is true for $Q_{v_i}$ and $\Delta_{v_i}$). 
Given this result, the magnitude of the correction term is notably small only for $x_p=30$. 
The statistics of the correction term indicate that the correction term is close to zero for $x_p=30$ (Fig.~\ref{fig:fig6}(b)).
On the basis of the results described in Secs.~\ref{sec:1d_TW} and \ref{sec:1d_OS}, our method successfully operated as intended for both target waves and oscillating spots.

\par 
As with the target waves (Sec.~\ref{sec:1d_TW}), ignoring the correction term changes the phase calculation results, and the extent of this change depends on the correction term.
Figure~\ref{fig:fig6}(c) shows the histograms of $\phi_1 - \phi_2$ calculated on the basis of Eqs.~(\ref{eq:1d_correction_term2}) and (\ref{eq:1d_phi}).
The histograms of $\phi_1-\phi_2$ for each $x_p$ are almost similar, although the histogram of $\theta_1^{x_p}-\theta_2^{x_p}$ shown in Fig.~\ref{fig:fig5}(b) varies depending on $x_p$.
It is evident that the histograms of $\theta_1^{x_p}-\theta_2^{x_p}$ and $\phi_1-\phi_2$  differ significantly for $x_p=10,20,40$ since the magnitude of the correction term is large, but they are similar for $x_p = 30$ since the correction term is nearly zero.
Ignoring the correction term degrades the accuracy of phase equation estimation unless the correction term is nearly zero.
Figures~\ref{fig:figA2}(c) and~\ref{fig:figA2}(d) in Appendix~\ref{appendix:estimation}, which show the phase equations estimated from the time series of $\theta_i^{x_p}(t)$ and $\phi_i(t)$, respectively. 
The phase equations estimated from $\theta_i^{x_p}$ differ significantly from the true forms for $x_p=10,20,40$, but closely match the true forms for $x_p=30$. 
In contrast, the phase equations estimated from $\phi_i$ are consistent with the true forms regardless of $x_p$.

%%%%%%%%%%%%%%%%%%%%%%%%%%%%%%%%%%%%%%%%%%%%%%%%%%%%%%%%%%%%%
%%%%%%%%%%%%%%%%%%%%%%%%%%%%%%%%%%%%%%%%%%%%%%%%%%%%%%%%%%%%%
%%%%%%%%%%%%%%%%%%%%%%%%%%%%%%%%%%%%%%%%%%%%%%%%%%%%%%%%%%%%%

\section{CALCULATING THE PHASE FROM MEASUREMENTS ON ALL SPATIAL GRID POINTS}
\label{sec:PCA}

\par
We explore a method for calculating the phase using the Poincar\'e section applied to time series obtained by the orthogonal decomposition of the spatiotemporal dynamics. 
The orthonormal basis functions used in this decomposition are assumed to be obtained through PCA.
We explain the orthogonal decomposition technique (Sec.~\ref{sec:PCA_decomposition}), the linear interpolation to calculate the phase (Sec.~\ref{sec:PCA_phasecal}), and the differences between the calculated phase and isochron-based phase (Sec.~\ref{sec:PCA_correction_term}).
Then, we illustrate the phase calculation for spatiotemporal dynamics using two different decomposition schemes (Sec.~\ref{sec:PCA_uv} and Sec.~\ref{sec:PCA_QuQv}).

\subsection{Orthogonal decomposition}
\label{sec:PCA_decomposition}

\par
Before delving into the specifics of the FHN reaction-diffusion model, we consider the general form of the reaction-diffusion model described by Eq.~(\ref{eq:RDmodel}).
The $n$th entries of the state vector $\bm{X}_i$, limit-cycle solution $\bm{\chi}_i$, and phase sensitivity function $\bm{Q}_i$ are represented as $X_{i,n}$, $\chi_{i,n}$, and $Q_{i,n}$, respectively.
For decomposition, we project them onto the orthonormal basis functions $\Phi_{i,n}^k$ ($k=1,2,\ldots, K$) as follows:
\begin{align}
    \label{eq:PCA_projection}
    \begin{split}
    & X_{i,n}^k(t) = \int_0^L X_{i,n}^k(x,t) \Phi_{i,n}^k(x) \mathrm{d}x, \\
    & \chi_{i,n}^k(\phi_i) = \int_0^L \chi_{i,n}^k(x,\phi_i) \Phi_{i,n}^k(x) \mathrm{d}x,\\
    & Q_{i,n}^k(\phi_i) = \int_0^L Q_{i,n}^k(x,\phi_i) \Phi_{i,n}^k(x) \mathrm{d}x,
    \end{split}
\end{align}
where the basis function satisfies $\int_0^L \Phi_{i,n}^p(x) \Phi_{i,n}^q(x) \mathrm{d}x= \delta_{pq}$ ($\delta_{pq}$ denotes the Kronecker delta).
According to Eq.~(\ref{eq:PCA_projection}), the infinite-dimensional state space is mapped to $KN$-dimensional state space.

\par
We rewrite Eq.~(\ref{eq:PCA_projection}) for the FHN reaction-diffusion model.
Given the orthonormal basis functions $\Phi_{u_i}^k(x)$ ($k=1,2,\ldots,K$), the projections of the state variable, limit cycle solution, and phase sensitivity function associated with $u_i$ onto the basis functions are calculated as follows:
\begin{align}
    \label{eq:PCA_projection2}
    \begin{split}
    & u_i^{k}(t) = \int_0^L u_i(x,t) \Phi_{u_i}^k(x) \mathrm{d}x, \\
    & \chi_{u_i}^{k}(\phi_i) = \int_0^L \chi_{u_i}(x,\phi_i) \Phi_{u_i}^k(x) \mathrm{d}x, \\
    & Q_{u_i}^{k}(\phi_i) = \int_0^L Q_{u_i}(x,\phi_i) \Phi_{u_i}^k(x) \mathrm{d}x,
    \end{split}
\end{align}
where the basis functions satisfies $\int_0^L \Phi_{u_i}^p(x) \Phi_{u_i}^q(x) \mathrm{d}x= \delta_{pq}$.
The variables $v_i^k(t)$, $\chi_{v_i}^k(\phi_i)$, and $Q_{v_i}^k(\phi_i)$ are similarly calculated using the orthonormal basis functions $\Phi_{v_i}^k(x)$ (the equations are abbreviated).
One scheme for obtaining these basis functions involves applying PCA to the time series of $u_i$ and $v_i$ collected from all spatial grid points (we describe this in Sec.~\ref{sec:PCA_uv}, and another approach is discussed in Sec.~\ref{sec:PCA_QuQv}). 
We set $K=50$ or $K=200$ to ensure that each function can be reproduced nearly $100\%$ on the basis of the following equations:
\begin{align}
    \begin{split}
    &u_i(x,t) \simeq \sum_{k=1}^K u_i^{k}(t) \Phi_{u_i}^k(x), \\
    & \chi_{u_i}(x,\phi_i) \simeq \sum_{k=1}^K \chi_{u_i}^{k}(\phi_i) \Phi_{u_i}^k(x), \\
    & Q_{u_i}(x,\phi_i) \simeq \sum_{k=1}^K Q_{u_i}^{k}(\phi_i) \Phi_{u_i}^k(x).
    \end{split}
\end{align}
The variables $v_i(x,t)$, $\chi_{v_i}(x,\phi_i)$, and $Q_{v_i}(x,\phi_i)$ are calculated in a similar way (the equations are abbreviated).

\subsection{Phase calculation by linear interpolation}
\label{sec:PCA_phasecal}

\par 
We calculate the phase $\theta_i^{k_p}$ from the time series of $u_i^{k_p}(x,t)$ or $v_i^{k_p}(x,t)$. 
The Poincar\'e section is applied to the time series of $u_i^{k_p}(t)$ or $v_i^{k_p}(t)$, recording the time $t_{i,j}^{k_p}$ when the time series intersects the Poincar\'e section for the $j$th time.
The phase $\theta_i^{k_p}$ is then calculated from the set of times, $\{t_{i,j}^{k_p}\}_j$, ensuring that it satisfies $\theta_{i,j}^{k_p}(t_{i,j}^{k_p})=2\pi j$.
As in Eq.~(\ref{eq:1d_theta}), we employ linear interpolation of the phase as follows:
\begin{align}
    \label{eq:PCA_theta}
    \theta_i^{k_p}(t)  
    = 2\pi j + 2\pi \frac{t-t_{i,j}^{k_p}}{t_{i,j+1}^{k_p}-t_{i,j}^{k_p}} 
    \hspace{3mm} (t_{i,j}^{k_p} \leq t \leq t_{i,j+1}^{k_p}). 
\end{align}
We assume $k_p=1$, and using another variable, $u_i^{k_p}(t)$ or $v_i^{k_p}(t)$ $(k_p \geq 2)$, is not discussed in this study.

\subsection{Difference between the calculated phase and isochron-based phase}
\label{sec:PCA_correction_term}

\par
We calculated the isochron-based phase $\phi_i(t)$, which obeys the isochron deformed by the orthogonal decomposition.
There is a difference between the phases, $\theta_i^{k_p}$ and $\phi_i$, which can be evaluated using a linear approximation in the vicinity of the limit-cycle solution.
Let us consider the general form of the reaction-diffusion model described in Eq.~(\ref{eq:RDmodel}). 
We assume that the Poincar\'e section is applied to the time series of $X_{i,n_p}^{k_p}$ and that the time $t_{i,j}^{k_p}$ is recorded when $X_{i,n_p}^{k_p}(t)$ intersects the Poincar\'e section for the $j$th time. 
The time series of the phase $\theta_{i}^{k_p}$ is calculated using Eq.~(\ref{eq:PCA_theta}).
We then define the correction term, $c_{i,j}^{k_p}$, as the difference between $\theta_i^{k_p}(t_{i,j}^{k_p})$ and $\phi_i(t_{i,j}^{k_p})$. 
The phase $\phi_i(t_{i,j}^{k_p})$ and correction term are calculated as follows:
\begin{align}
    \label{eq:PCA_correction_term}
    \begin{split}
    & \phi_i(t_{i,j}^{k_p})
    = \theta_i^{k_p}(t_{i,j}^{k_p}) + c_{i,j}^{k_p} 
    = 2 \pi j + c_{i,j}^{k_p},  \\
    & c_{i,j}^{k_p} = \sum_{(k, n) \neq (k_p, n_p)}^{(K,N)} Q_{i,n}^{k} (0) \Delta_{i,n}^k (t_{i,j}^{k_p}), \\
    & \Delta_{i,n}^k (t_{i,j}^{k_p}) := X_{i,n}^k(t_{i,j}^{k_p}) - \chi_{i,n}^k(0),
    \end{split}
\end{align}
where $\Delta_{i,n}^k$ is the deviations of $X_{i,n}^k$ from the state of $\phi_i=0$ on the limit-cycle $\chi_{i,n}^k$.
The entry of $(k_p, n_p)$ are removed from the summation in Eq.~(\ref{eq:PCA_correction_term}) since the two values, $X_{i,n_p}^{k_p}(t_{i,j}^{k_p})$ and $\chi_{i,n_p}^{k_p}(0)$, are equal on the Poincar\'e section, i.e., $\Delta_{i,n_p}^{k_p}(t_{i,j}^{k_p}) = 0$.
The derivation of Eq.~(\ref{eq:PCA_correction_term}) is detailed in Appendix~\ref{appendix:proof}.
Fig.~\ref{fig:fig7} illustrates that the two states in the $KN$-dimensional space, $\{\chi_{i,n}^k(0)\}_{n,k}$ and $\{X_{i,n}^k(t_{i,j}^{k_p})\}_{n,k}$, lie on the same Poincar\'e section $P$ but on different isochrons, $I(0)$ and $I(c_{i,j}^{k_p})$.
Note that the isochron is mapped from the infinite-dimensional space to the $KN$-dimensional space through orthogonal decomposition.
Unless the Poincar\'e section is the same as $I(0)$, the two states are located on distinct isochrons. 
The correction term represents the difference between the phases associated with these two isochrons.

\par 
We rewrite Eq.~(\ref{eq:PCA_correction_term}) for the FHN reaction-diffusion model. 
Here, we use the index $n_p \in \{u, v\}$ to denote whether the Poincar\'e section is applied to the time series of $u_i^1$ or $v_i^1$, instead of $1 \leq n_p \leq N$ used in Eq.~(\ref{eq:PCA_correction_term}).
When we apply the Poincar\'e section to the time series of $u_i^{k_p}$, i.e., $n_p=u$, the phase $\phi_i(t_{i,j}^{k_p})$ and correction term are calculated as follows:
\begin{align}
    \label{eq:PCA_correction_term2}
    \begin{split}
    & \phi_i(t_{i,j}^{k_p}) = 2 \pi j + c_{i,j}^{k_p},  \\
    & c_{i,j}^{k_p} =
    \sum_{k \neq k_p}^K Q_{u_i}^k (0) \Delta_{u_i}^k (t_{i,j}^{k_p})
    + \sum_{k=1}^K Q_{v_i}^k (0) \Delta_{v_i}^k (t_{i,j}^{k_p}),
    \end{split}
\end{align}
where $\Delta_{u_i}^k (t_{i,j}^{k_p}):=u_i^k(t_{i,j}^{k_p}) - \chi_{u_i}^k(0)$ and $\Delta_{v_i}^k (t_{i,j}^{k_p}):=v_i^k(t_{i,j}^{k_p}) - \chi_{v_i}^k(0)$ are the deviations of $u_i^k$ and $v_i^k$ from the state of $\phi_i=0$ on the limit-cycle solution.
When the Poincar\'e section is applied to the time series of $v_i^{k_p}$, i.e., $n_p=v$, the correction term is calculated as follows instead of Eq.~(\ref{eq:PCA_correction_term2}):
\begin{align}
    \label{eq:PCA_correction_term2'}
    & c_{i,j}^{k_p} =
    \sum_{k=1}^K Q_{u_i}^k (0) \Delta_{u_i}^k (t_{i,j}^{k_p}) 
    + \sum_{k \neq k_p}^K Q_{v_i}^k (0) \Delta_{v_i}^k (t_{i,j}^{k_p}). \tag{\ref{eq:PCA_correction_term2}'} 
\end{align}
Hereafter, when we refer to Eq.~(\ref{eq:PCA_correction_term2}), Eq.~(\ref{eq:PCA_correction_term2'}) is also included to the reference implicitly.
Then, we linearly interpolate the phase as follows:
\begin{align}
    \label{eq:PCA_phi}
    \hspace{-2mm}
    \phi_i(t) 
        &=  \phi_i(t_{i,j}^{k_p}) + 
        (\phi_i(t_{i,j+1}^{k_p})-\phi_i(t_{i,j}^{k_p})) \frac{t-t_{i,j}^{k_p}}{t_{i,j+1}^{k_p}-t_{i,j}^{k_p}} \nonumber \\
        & = 2 \pi j + c_{i,j}^{k_p}+ (2\pi + c_{i,j+1}^{k_p}-c_{i,j}^{k_p}) \frac{t-t_{i,j}^{k_p}}{t_{i,j+1}^{k_p}-t_{i,j}^{k_p}} \\
        & (t_{i,j}^{k_p} \leq t \leq t_{i,j+1}^{k_p}) \nonumber.
\end{align}
The transformation to the second row is achieved by substituting Eq.~(\ref{eq:PCA_correction_term2}) into the first row. 
According to Eq.~(\ref{eq:PCA_correction_term}) or Eq.~(\ref{eq:PCA_correction_term2}), the magnitude of the correction term depends on both the deviation magnitude from the limit-cycle solution and the amplitude of the (decomposed) phase sensitivity functions.
Therefore, when $u_i^k$ or $v_i^k$ fluctuates significantly and the corresponding $Q_{u_i}^k$ or $Q_{v_i}^k$ has a large amplitude, the magnitude of the correction term becomes large.
We present the results of phase calculations for both $n_p=u$ and $n_p=v$, but selecting $n_p$ does not inherently improve the accuracy of the phase calculation.

\subsection{Case of using the basis function obtained from the spatiotemporal dynamics}
\label{sec:PCA_uv}

\par
We illustrate the process of calculating phases for spatiotemporal dynamics that display target waves and oscillating spots. 
In this context, we assume that the basis functions are obtained by applying PCA to the spatiotemporal dynamics of $u_i(x,t)$ and $v_i(x,t)$, respectively.
The index $k=1,2,\ldots,K$ is arranged in descending order based on the contribution ratio.

\par 
To begin, we calculate the phase of spatiotemporal dynamics exhibiting target waves.
Fig.~\ref{fig:fig8}(a) shows the amplitudes of $Q_{u_i}^k$ and $Q_{v_i}^k$, revealing that many of these amplitudes are not nearly zero. 
Therefore, fluctuations in the corresponding $u_i^k$ and $v_i^k$ influence the correction term. 
We applied the Poincar\'e section to either time series of $u_i^{k_p}$ or $v_i^{k_p}$~($k_p=1$) to calculate $\theta_i^{k_p}$ as shown in Fig.~\ref{fig:fig8}(b). 
The histograms of $\theta_1^{k_p}(t)-\theta_2^{k_p}$ for $n_p=u$ and $n_p=v$ shown in Fig.~\ref{fig:fig8}(c) slightly differ from the distribution of $\phi_1 - \phi_2$ calculated from the phase equations.
The statistics of the correction terms are shown in Fig.~\ref{fig:fig8}(d). 
It indicates that the magnitudes of the correction term for both $n_p=u$ and $n_p=v$ are much smaller compared to those calculated from $u_i(x_p, t)$ for $x_p \simeq 15, 30$ (Fig.~\ref{fig:fig4}(b)).
Although the magnitude of the correction term for $n_p=u$ is smaller compared to that for $n_p=v$, the difference between them is not a primary concern in this study.
To confirm the correction term certainly represents the difference between $\theta_i^{k_p}(t_{i,j}^{k_p})$ and $\phi_i(t_{i,j}^{k_p})$ correctly, we calculated the histogram of $\phi_1-\phi_2$ on the basis of Eqs.~(\ref{eq:PCA_correction_term2}) and (\ref{eq:PCA_phi}). 
Figure~\ref{fig:fig8}(e) indicates that both histograms of $\phi_1-\phi_2$ are more similar to the distribution calculated from the phase equations than the histograms of $\theta_1^{k_p}-\theta_2^{k_p}$ shown in Fig.~\ref{fig:fig8}(c). 
These results support the validity of the correction terms.
Furthermore, Figs.~\ref{fig:figA3}(a) and ~\ref{fig:figA3}(b) in Appendix~\ref{appendix:estimation} show the phase equations estimated from the time series of $\theta_i^{k_p}$ and $\phi_i$, respectively.
The phase equations estimated from the time series of $\theta_i^{k_p}$ resemble the true forms qualitatively, while those estimated from the time series of $\phi_i$ appear to match the true forms even more closely.

\par 
Next, we calculated the phases of the spatiotemporal dynamics exhibiting oscillating spots.
Since the amplitude of $Q_{u_i}^k$ and $Q_{v_i}^k$ up to about $k=5$ are nonzero (Fig.~\ref{fig:fig9}(a)), the fluctuations in the corresponding $u_i^k$ and $v_i^k$ influence the correction term.
We applied the Poincar\'e section to either time series of $u_i^{k_p}$ or $v_i^{k_p}$ ($k_p=1$) to calculate $\theta_i^{k_p}$ as shown in Fig.~\ref{fig:fig9}(b).
The histograms of $\theta_1^{k_p}(t)-\theta_2^{k_p}$ shown in Fig.~\ref{fig:fig9}(c) differs from the distribution of $\phi_1 - \phi_2$ calculated from the phase equations.
The large difference between the histogram and distribution is supported by the statistics of the correction term shown in Fig.~\ref{fig:fig9}(d).
The statistics indicate that the magnitude of the correction term is comparable to those calculated from $u_i(x_p, t)$ for $x_p=10, 20, 40$ (Fig.~\ref{fig:fig6}(b)).
Figure~\ref{fig:fig9}(e) indicates that both histograms of $\phi_1-\phi_2$, which are calculated from the time series of isochron-based phases, are similar to the distribution calculated from the phase equations. 
These findings validate that the correction term represents the difference between $\theta_i^{k_p}(t_{i,j}^{k_p})$ and $\phi_i(t_{i,j}^{k_p})$. 
Furthermore, Figs.~\ref{fig:figA3}(c) and ~\ref{fig:figA3}(d) in Appendix~\ref{appendix:estimation} show the phase equations estimated from the time series of $\theta_i^{k_p}$ and $\phi_i$, respectively. 
The phase equations estimated from $\theta_i^{k_p}$ deviate from the true forms while those estimated from $\phi_i$ closely match the true forms.

\subsection{Case of using the basis function obtained from the phase sensitivity function}
\label{sec:PCA_QuQv}

\par
According to Eqs.~(\ref{eq:PCA_projection2}) and (\ref{eq:PCA_correction_term2}), the correction term can vary depending on the choice of basis functions.
Fluctuations in $u_i^k$ and $v_i^k$ do not influence the correction term when their corresponding $Q_{u_i}^k$ and $Q_{v_i}^k$ values are nearly zero.
Therefore, the magnitude of the correction term is expected to be small if the amplitudes of $Q_{u_i}^k$ and $Q_{v_i}^k$ are localized to just a few components.
To achieve this localization, we propose using basis functions obtained by applying PCA to the phase sensitivity function, rather than the spatiotemporal dynamics.
This method ensures that the basis functions are determined to maximize the representation of the phase sensitivity functions with as few $Q_{u_i}^k$ and $Q_{v_i}^k$ as possible. 
The index $k = 1,2,...,K$ is arranged in descending order based on their contribution ratio related to the phase sensitivity function.

\par
First, we examine the phase calculation for spatiotemporal dynamics exhibiting target waves. 
Figure~\ref{fig:fig10}(a) indicates that the amplitude of $Q_{u_i}^k$ and $Q_{v_i}^k$ are localized to $k=1$.
In such a situation, the term that contains $v_i^1$ mainly determines the correction term in the case of $n_p=u$, while the term that contains $u_i^1$ does the same in the case of $n_p=v$ (Eq.~(\ref{eq:PCA_correction_term2})). 
This indicates that the magnitude of the correction term is primarily determined by the fluctuations in either $u_i^1$ or $v_i^1$.
Since the trajectory of $(u_1^1, v_1^1)$ is close to the limit-cycle of $(\chi_{u_1}^1, \chi_{v_i}^1)$ as shown in Fig.~\ref{fig:fig10}(b), the magnitude of the correction terms is expected to be relatively small. 
We applied the Poincar\'e section to the time series of $u_i^{k_p}$ or $v_i^{k_p}$ ($k_p=1$) to calculate $\theta_i^{k_p}$ (Fig.~\ref{fig:fig10}(c)).
The histograms of $\theta_1^{k_p}-\theta_2^{k_p}$ shown in Fig.~\ref{fig:fig10}(d) slightly differ from the distribution of $\phi_1-\phi_2$ calculated from the phase equations.
Furthermore, according to the statistics of the correction term shown in Fig.~\ref{fig:fig10}(e), its magnitude is much smaller compared to those shown in  Fig.~\ref{fig:fig8}(d) for both $n_p=u$ and $n_p=v$. 
These results indicate that using the basis functions obtained by applying PCA to the phase sensitivity function reduces the magnitude of the correction term.
Since the correction term is nearly negligible, ignoring it does not substantially degrade the accuracy of phase equation estimation.
Figure~\ref{fig:figA4}(a) in Appendix~\ref{appendix:estimation} shows that the phase equations estimated from the time series of $\theta_i^{k_p}$ closely resemble the true forms.

\par
Next, we examine the phase calculation for spatiotemporal dynamics exhibiting oscillating spots. 
Figure~\ref{fig:fig11}(a) indicates that $Q_{u_i}^k$ and $Q_{v_i}^k$ up to about $k=5$ have nonzero amplitude. 
Therefore, fluctuations in $u_i^k$ and $v_i^k$ up to about $k=5$ determine the correction terms.
Figure~\ref{fig:fig11}(b) shows the deviation of the trajectory of $(u_1^1, v_1^1)$ from the limit-cycle of $(\chi_{u_1}^1,\chi_{v_i}^1)$ is larger compared to the target waves shown in Fig.~\ref{fig:fig10}(b).
Thus, the magnitude of the correction term is likely to be larger for oscillating spots than for target waves. 
We applied the Poincar\'e section to the time series of $u_i^{k_p}$ or $v_i^{k_p}$ ($k_p=1$) to calculate $\theta_i^{k_p}$ (Fig.~\ref{fig:fig11}(c)).
The histograms of $\theta_1^{k_p}-\theta_2^{k_p}$ differ from the distribution of $\phi_1-\phi_2$ calculated from the phase equations (Fig.~\ref{fig:fig11}(d)).
Furthermore, the statistics of the correction term shown in Fig.~\ref{fig:fig11}(e) indicate that its magnitude is comparable to that shown in Fig.~\ref{fig:fig9}(d). 
These results indicate that this PCA-based scheme does not reduce the magnitude of the correction term.
Furthermore, ignoring the correction term degrades the accuracy of the phase equation estimation.
Figure~\ref{fig:figA4}(b) in Appendix~\ref{appendix:estimation} shows that the phase equations estimated from the time series of $\theta_i^{k_p}$ differ from the true forms.

\par 
In this paragraph, we describe why applying PCA to the phase sensitivity function effectively reduced the correction term for target waves but was less successful for oscillating spots.
As previously discussed, the localization of the amplitudes of $Q_{u_i}^k$ and $Q_{v_i}^k$ is an important factor. 
According to Eq.~(\ref{eq:PCA_correction_term2}), when $Q_{u_i}^k$ and $Q_{v_i}^k$ have large amplitudes, fluctuations in $u_i^k$ and $v_i^k$ significantly influence the correction term.
Thus, the degree of localization is related to the extent to which the correction term can be reduced.
For target waves, where the phase sensitivity function exhibits an oscillatory pattern (see Fig.~\ref{fig:figA1}(b) in Appendix~\ref{appendix:limitcycle}), $Q_{u_i}^1$ and $Q_{v_i}^1$ predominantly capture the phase sensitivity function.
By contrast, in the case of oscillating spots where the phase sensitivity function exhibits the translational pattern (see Fig.~\ref{fig:figA1}(d) in Appendix~\ref{appendix:limitcycle}), several $Q_{u_i}^k$ and $Q_{v_i}^k$ are necessary to approximate it.
In summary, the PCA-based scheme where the PCA is applied to the phase sensitivity function is effective for target waves since only fluctuation of either $u_i^1$ or $v_i^1$ impacts the correction term. 
By contrast, this scheme is less effective for oscillating spots, where the fluctuations of several $u_i^k$ and $v_i^k$ influence the correction term.
Furthermore, the magnitude of these fluctuations plays a crucial role in determining the magnitude of the correction term.
For example, given that $Q_{u_i}^1$ and $Q_{v_i}^1$ have the largest amplitude, the fluctuation in either $u_i^1$ or $v_i^1$ remains a significant contributor to the correction term (see Eq.~(\ref{eq:PCA_correction_term2})).

%%%%%%%%%%%%%%%%%%%%%%%%%%%%%%%%%%%%%%%%%%%%%%%%%%%%%%%%%%%%%
%%%%%%%%%%%%%%%%%%%%%%%%%%%%%%%%%%%%%%%%%%%%%%%%%%%%%%%%%%%%%
%%%%%%%%%%%%%%%%%%%%%%%%%%%%%%%%%%%%%%%%%%%%%%%%%%%%%%%%%%%%%

\section{DISCUSSION}
\label{sec:discussion}

\par 
In this study, we investigated methods to accurately calculate the phase time series by applying the Poincar\'e section to one-dimensional time series.
We developed the approaches to reduce the difference between the calculated phase and the isochron-based phase due to the discrepancy between the Poincar\'e section and the isochron.
In this study, this difference was referred to as the correction term.
We utilized the weakly coupled FHN reaction-diffusion models (Sec.~\ref{sec:model_FHNRD}) exhibiting target waves (Sec.~\ref{sec:model_TW}) and oscillating spots (Sec.~\ref{sec:model_OS}) for the numerical simulations.
In Sec~\ref{sec:1d}, we examined the approach for calculating the phase from measurements taken at a single spatial grid point. 
We detailed the linear interpolation method for phase calculation using the Poincar\'e section (Sec.~\ref{sec:1d_phasecal}) and explained our approach for reducing the magnitude of the correction term (Sec.~\ref{sec:1d_position}).
We demonstrated this approach for spatiotemporal dynamics exhibiting target waves and oscillating spots (Secs.~\ref{sec:1d_TW} and \ref{sec:1d_OS}).
In Sec~\ref{sec:PCA}, we explored the method for calculating the phase time series from measurements taken across all spatial grid points. 
We describe the orthogonal decomposition of spatiotemporal dynamics, the limit-cycle solution, and the phase sensitivity function (Sec.~\ref{sec:PCA_decomposition}). 
Following this, we explain the linear interpolation method for phase calculation using the Poincar\'e section (Sec.~\ref{sec:PCA_phasecal}).
We proved that the two factors influence the magnitude of the correction terms (Sec.~\ref{sec:PCA_correction_term}). 
Finally, we demonstrated the phase calculation using two different schemes (Secs.~\ref{sec:PCA_uv} and \ref{sec:PCA_QuQv}).

\par 
In Sec.~\ref{sec:1d}, we focused on phase calculation from measurements at a single spatial grid point.
We found that the magnitude of the correction term is small when we apply the Poincar\'e section to the time series of measurements from a single grid point that belongs to the pacemaker region emitting target waves or the region where the spot's front exists.
The correction term is influenced by two factors: the spatially non-uniform phase sensitivity function and the deviation of the spatiotemporal dynamics from the limit-cycle solution (see Eq.~(\ref{eq:1d_correction_term})).
Specifically, when regions with a large amplitude of the phase sensitivity function overlap those having a large deviation, the magnitude of the correction term will be large.
Thus, to avoid overlapping, we developed an approach to calculate the phase from measurements taken in regions with a large amplitude of the phase sensitivity function.
We applied this approach to the rhythmic patterns of target waves and oscillating spots.
These patterns are known to have regions, such as the pacemaker region or the spot’s front, where the phase sensitivity function has a large amplitude.
Although the phase sensitivity function for real-world systems is generally unknown, it can be inferred that it is localized in regions controlling the overall dynamics~\cite{Nakao_2014}.
Therefore, using insights from experimental and theoretical studies to determine the optimal measurement positions is advisable.

\par 
In Sec.~\ref{sec:PCA}, we focused on calculating the phase from one-dimensional time series obtained from measurements across all spatial grid points using orthogonal decomposition.
We proved that the magnitude of the correction term is influenced by two factors: the magnitude of fluctuation in the (decomposed) spatiotemporal dynamics and the amplitude of the (decomposed) phase sensitivity functions (see Eq.~(\ref{eq:PCA_correction_term})).
On the basis of this finding, we proposed a scheme where PCA is applied to the phase sensitivity function rather than the spatiotemporal dynamics in Sec.~\ref{sec:PCA_QuQv}.
This scheme aims to localize the amplitude of the (decomposed) phase sensitivity function to just a few components.
The results show that for the target-wave solution, the amplitude of the (decomposed) phase sensitivity function was localized to a single component, whereas for the oscillating spot solution, it was not.
This difference between the two solutions arises from whether the phase sensitivity function exhibits an oscillatory pattern or translational pattern. 
The latter case is difficult to decompose to a few components using PCA.
Furthermore, we confirmed that the fluctuations in the (decomposed) spatiotemporal dynamics are small for target waves but not for oscillating spots.
Consequently, we concluded that applying PCA to the phase sensitivity function can improve the accuracy of phase calculation for target waves.
It is worth noting that phase calculation for target waves can still achieve a certain degree of accuracy even without this PCA scheme (see Sec.~\ref{sec:PCA_uv}).

\par 
Properly setting the Poincar\'e section is crucial not only for continuous systems but also for real-world discrete systems, including the network of dynamical elements exhibiting the collective oscillation~\cite{Kori_2009, Kawamura_2015, Nakao_2018}.
This necessity arises because a large magnitude of the correction term can also occur in high-dimensional ODE systems, especially when the dimension of the dynamical system is comparable to the number of spatial grid points in PDE cases. 
This can be explained by the fact that the expressions for the correction terms in the ODE and PDE cases have a similar form. (see Eq.~(\ref{eq:ODEperturbation}) in Appendix~\ref{appendix:isochron_ODE} and Eq.~(\ref{eq:perturbation2}) in Appendix~\ref{appendix:isochron}).
Furthermore, although we assumed the one-dimensional space, the approach for setting the Poincar\'e section discussed in this study is also applicable to multidimensional spaces. 
This is because the localization of phase sensitivity functions for rhythmic patterns occurs similarly in multidimensional spaces~\cite{Nakao_2014}.

\par 
The findings of this study provide insights for calculating the phase from measurements across various domains such as meteorology~\cite{Read_2012, Rajendran_2016}, electrochemical oscillators~\cite{Kiss_2002}, biophysics~\cite{Nakagaki_2000}, and life science~\cite{Guevara_1981}.
When measuring at a single grid point, the accuracy of phase calculation is improved by taking measurements from the pacemaker region or the region where the front of the oscillating spot exists. 
In addition, for systems exhibiting target waves, it is possible to calculate the phase from a one-dimensional time series obtained by applying PCA to multidimensional measurements over multiple spatial grid points.
For example, in meteorology, PCA (also known as empirical orthogonal functions) is sometimes applied to spatiotemporal dynamics to obtain the time series of modes (and its basis function). 
Our study suggests that these modes can sometimes be used to calculate the phase. 
The findings of this study can be experimentally verified using a pair of photosensitive Belousov-Zhabotinsky systems, where two spatiotemporal rhythmic patterns are locally coupled via video cameras and projectors~\cite{Hildebrand_synchronization_2003}.

\par 
Our key finding allows for the calculation of the phase of collective oscillations of dynamical elements ~\cite{Kori_2009, Kawamura_2015, Nakao_2018} and spatiotemporal dynamics~\cite{Nakao_2014} from measurements.
Therefore, the phase response, phase sensitivity function, and phase coupling function of the collective oscillation and spatiotemporal dynamics can be estimated by combining our method with conventional estimation methods~\cite{Rosenblum_2001,  Galan_2005, Miyazaki_2006, Tokuda_2007, Ermentrout_2007, Kralemann_2007, Kralemann_2008, Ota_2009, Stankovski_2012, Duggento_2012, Kralemann_2013, Ota_2014, Imai_2017, Stankovski_2017, Cestnik_2017, Cestnik_2018} as we did in Appendix~\ref{appendix:estimation}.
To ensure accurate estimation, it is crucial to reduce the magnitude of the correction term caused by the discrepancy between the Poincar\'e section and the isochron.
This study revealed a proper setting for the Poincar\'e section that realizes the calculation of the phase with a small magnitude of the correction term.
The setting is based on the spatial localization of the phase sensitivity function.
Ignoring these considerations can lead to incorrect estimations and misunderstandings of system properties, as shown by the incorrectly estimated phase equations (see Appendix~\ref{appendix:estimation}).

\par 
This study showed that even in the absence of noise, a difference exists between the calculated phase and the isochron-based phase, and we investigated methods to reduce this difference.
Addressing the robustness of phase calculation under the noise remains a future task.
Furthermore, in this study, we assumed that the basis functions for the orthogonal decomposition were obtained using PCA.
Other techniques, such as extended dynamic mode decomposition (EDMD)~\cite{Williams_2015, Korda_2018}, which identifies not self-adjoint left and right vectors from multidimensional measurements, could also be used for orthogonal decomposition.
Developing a method that utilizes EDMD is a future challenge and might enable the calculation of the phase for spatiotemporal dynamics exhibiting oscillating spots with a smaller magnitude of the correction term.
Finally, some spatiotemporal dynamics have multiple phases, e.g., the governing equation has a limit-torus solution~\cite{Kawamura_2019}. 
Future research should also address the phase calculation for the systems exhibiting multiple rhythms.

%%%%%%%%%%%%%%%%%%%%%%%%%%%%%%%%%%%%%%%%%%%%%%%%%%%%%%%%%%%%%
%%%%%%%%%%%%%%%%%%%%%%%%%%%%%%%%%%%%%%%%%%%%%%%%%%%%%%%%%%%%%
%%%%%%%%%%%%%%%%%%%%%%%%%%%%%%%%%%%%%%%%%%%%%%%%%%%%%%%%%%%%%

%%

\begin{figure*}[h]
    \begin{center}
    \includegraphics[scale=1.0]{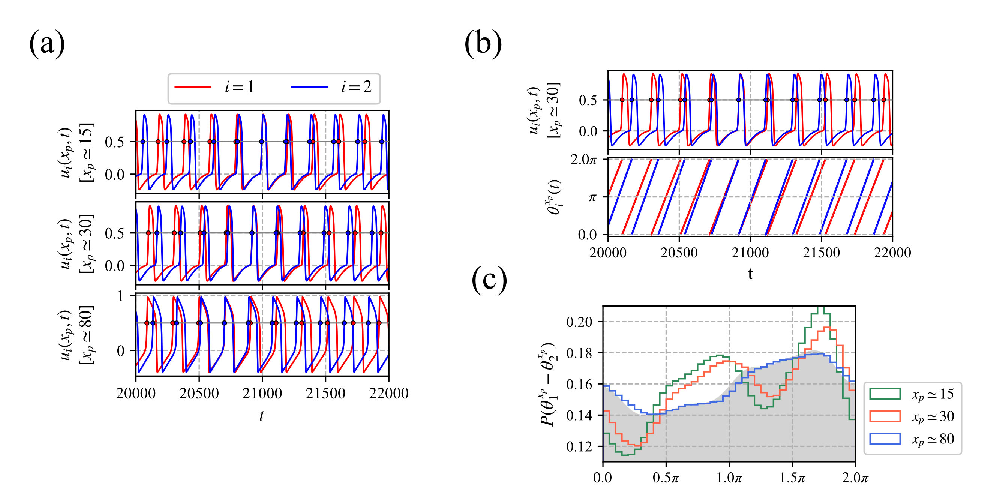}
    \caption{
    Time series of $u_i(x_p, t)$ and the calculation of $\theta_i^{x_p}(t)$ for the spatiotemporal dynamics exhibiting target waves.
    (a)
    Time series of $u_i(x_p, t)$ and the Poincar\'e sections for $x_p \simeq 15, 30, 80$.
    The red and blue lines indicate $i=1$ and $i=2$, respectively.
    The horizontal lines depict the Poincar\'e sections, and the dots represent the times at which $u_i(x_p, t)$ intersects the Poincar\'e section from negative to positive.
    (b)
    Example of the calculation of $\theta_i^{x_p}(t)$ for $x_p \simeq 30$. 
    The red and blue colors indicate $i=1$ and $i=2$, respectively. 
    (top) Recording the time of intersection.
    The dots represent the time $t_{i,j}^{x_p}$.
    (bottom) Calculation of the time series of $\theta_i^{x_p}(t)$ using the linear interpolation (see Eq.~(\ref{eq:1d_theta})). 
    The phase increases linearly by $2\pi$ during $t_{i,j+1}^{x_p}-t_{i,j}^{x_p}$.
    (c)
    Histograms of $\theta_1^{x_p}-\theta_2^{x_p}$ for $x_p \simeq 15$ (green), $30$ (red), and $80$ (blue).
    The histograms are calculated over a duration in which $|\theta_1^{x_p}-\theta_2^{x_p}|$ increases by $200\pi$.
    The distribution calculated from the phase equations is displayed in gray.
    }
    \label{fig:fig2}
    \end{center}
\end{figure*}

\begin{figure*}[h]
    \begin{center}
    \includegraphics[scale=1.0]{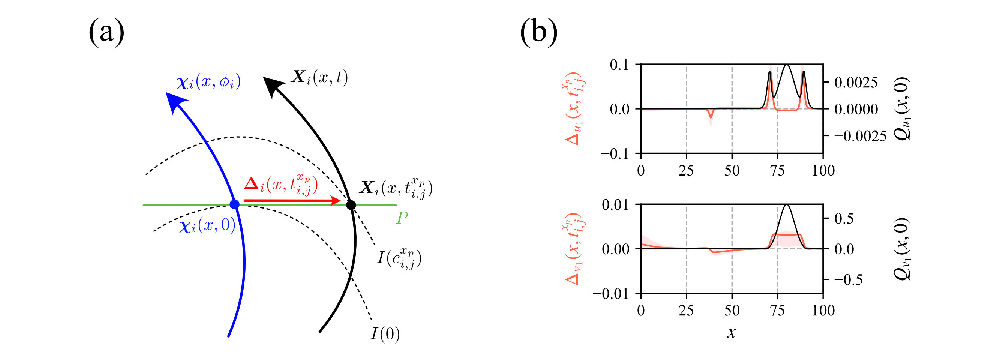}
    \caption{
    Illustration of calculation of the correction terms.
    (a)
    Illustration of two states on the same Poincar\'e section.
    The two states, $\bm{\chi}_i(x, 0)$ and $\bm{X}_i(x,t_{i,j}^{x_p})$, lie on the same Poincar\'e section $P$ but on different isochrons, $I(0)$ and $I(c_{i,j}^{x_p})$, respectively.
    The difference between these states is represented as $\bm{\Delta}_i(x,t_{i,j}^{x_p})$, whose $n$th entry is denoted as $\Delta_{i,n}$ used in Eq.(\ref{eq:1d_correction_term}).
    These states are located on distinct isochrons unless the Poincar\'e section is the same as $I(0)$. 
    The correction term quantifies the difference between the phases assigned to these distinct isochrons. 
    (b)
    Distributions of $\Delta_{u_1}(x,t_{1,j}^{x_p})$ and $\Delta_{v_1}(x,t_{1,j}^{x_p})$, along with the phase sensitivity functions, $Q_{u_1}(x,0)$ and $Q_{v_1}(x,0)$, for the spatiotemporal dynamics exhibiting target waves under the condition of $x_p \simeq 30$.
    The top figure displays $\Delta_{u_1}(x,t_{1,j}^{x_p})$ and $Q_{u_1}(x,0)$, and
    the bottom figure displays $\Delta_{v_1}(x,t_{1,j})$ and $Q_{v_1}(x,0)$.
    The red lines indicate the medians of $\Delta_{u_1}(x,t_{1,j}^{x_p})$ and $\Delta_{v_1}(x,t_{1,j}^{x_p})$.
    The red shades indicate the range between the $25$th and $75$th percentiles.
    The black lines indicate $Q_{u_1}(x,0)$ and $Q_{v_1}(x,0)$. 
    }
    \label{fig:fig3}
    \end{center}
\end{figure*}

\begin{figure*}[h]
    \begin{center}
    \includegraphics[scale=1.0]{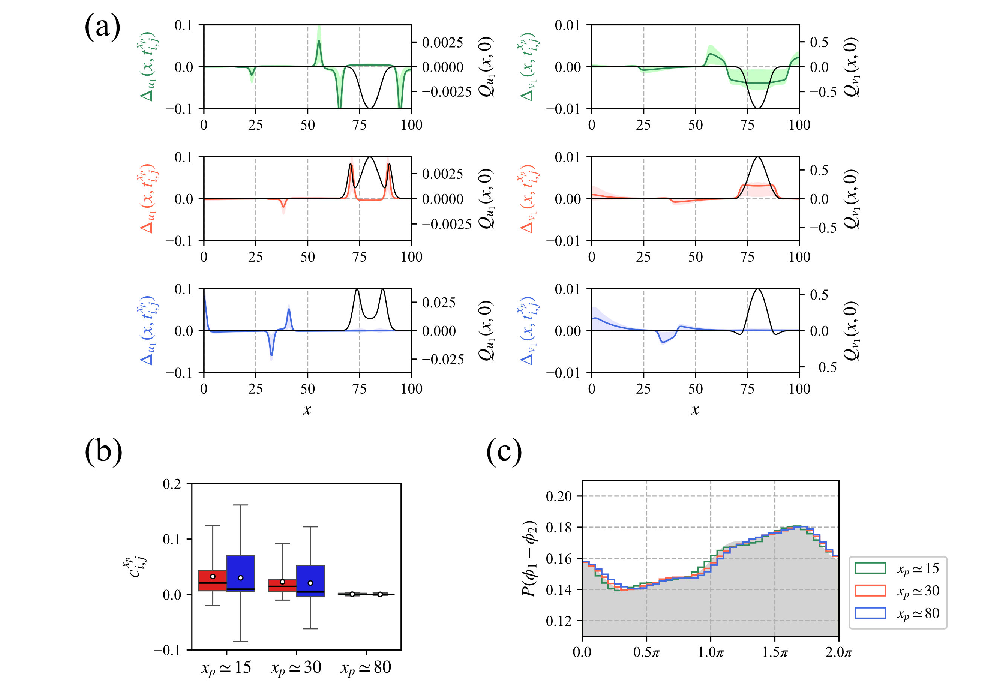}
    \caption{
    Calculation of the correction terms for spatiotemporal dynamics exhibiting target waves.
    (a)
    Distribution of $\Delta_{u_1}(x,t_{1,j}^{x_p})$ and $\Delta_{v_1}(x,t_{1,j}^{x_p})$, along with the phase sensitivity functions, $Q_{u_1}(x,0)$ and $Q_{v_1}(x,0)$, for the spatiotemporal dynamics exhibiting target waves under the condition of $x_p \simeq 15$ (top), $30$ (middle), and $80$ (bottom).
    The left figures display $\Delta_{u_1}(x,t_{1,j}^{x_p})$ and $Q_{u_1}(x,0)$, and the right figures display $\Delta_{v_1}(x,t_{1,j})$ and $Q_{v_1}(x,0)$.
    The colored lines indicate the medians of $\Delta_{u_1}(x,t_{1,j}^{x_p})$ and $\Delta_{v_1}(x,t_{1,j}^{x_p})$.
    The colored shades indicate the range between the $25$th and $75$th percentiles.
    The black lines indicate $Q_{u_1}(x,0)$ and $Q_{v_1}(x,0)$. 
    (b)
    Statistics of the correction term $c_{i,j}^{x_p}$ for $x_p \simeq 15, 30, 80$.
    The red ($i=1$) and blue ($i=2$) boxes represent the 1st and 3rd quartiles of dataset $\{c_{i,j}^{x_p}\}_j$. 
    The horizontal lines mark the medians, while the dots mark the averages. 
    The whiskers extend to show the maximum and minimum values.
    (c)
    Histograms of $\phi_1-\phi_2$ for $x_p \simeq 15$ (green), $30$ (red), and $80$ (blue).
    The histograms are calculated over a duration in which $|\phi_1-\phi_2|$ increases by $200\pi$.
    The distribution calculated from the phase equations is displayed in gray.
    }
    \label{fig:fig4}
    \end{center}
\end{figure*}

\begin{figure}[h]
    \begin{center}
    \includegraphics[scale=1.0]{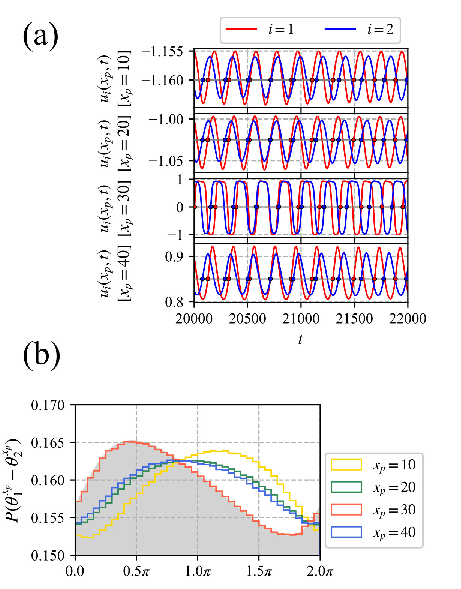}
    \caption{
    Time series of $u_i(x_p, t)$ and the calculation of $\theta_i^{x_p}(t)$ for the spatiotemporal dynamics exhibiting oscillating spots.
    (a)
    Time series of $u_i(x_p, t)$ and the Poincar\'e sections for $x_p = 10, 20, 30, 40$.
    The red and blue lines indicate $i=1$ and $i=2$, respectively.
    The horizontal lines depict the Poincar\'e sections, while the dots represent the times at which $u_i(x_p, t)$ intersects the Poincar\'e section from negative to positive.
    (b)
    Histograms of $\theta_1^{x_p}-\theta_2^{x_p}$ for $x_p = 10$ (yellow), $20$ (green), $30$ (red), and $40$ (blue).
    The histograms are calculated over a duration in which $|\theta_1^{x_p}-\theta_2^{x_p}|$ increases by $200\pi$.
    The distribution calculated from the phase equations is displayed in gray.
    }
    \label{fig:fig5}
    \end{center}
\end{figure}

\begin{figure*}[h]
    \begin{center}
    \includegraphics[scale=1.0]{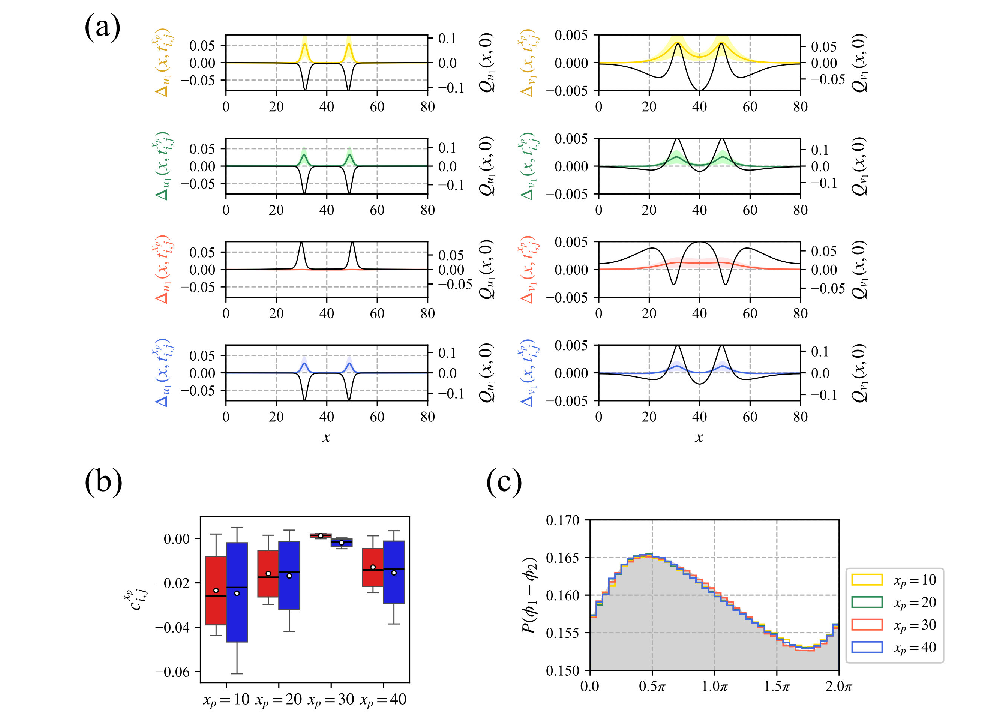}
    \caption{
    Calculation of the correction terms for the spatiotemporal dynamics exhibiting oscillating spots.
    (a)
    Distribution of $\Delta_{u_1}(x,t_{1,j}^{x_p})$ and $\Delta_{v_1}(x,t_{1,j}^{x_p})$, along with the phase sensitivity functions, $Q_{u_1}(x,0)$ and $Q_{v_1}(x,0)$, for the spatiotemporal dynamics exhibiting target waves under the condition of $x_p = 10$ (1st row), $20$ (2nd row), $30$ (3rd row), and $40$ (4th row).
    The left figures display $\Delta_{u_1}(x,t_{1,j}^{x_p})$ and $Q_{u_1}(x,0)$, and the right figures display $\Delta_{v_1}(x,t_{1,j})$ and $Q_{v_1}(x,0)$.
    The colored lines indicate the medians of $\Delta_{u_1}(x,t_{1,j}^{x_p})$ and $\Delta_{v_1}(x,t_{1,j}^{x_p})$.
    The colored shades indicate the range between the $25$th and $75$th percentiles.
    The black lines indicate $Q_{u_1}(x,0)$ and $Q_{v_1}(x,0)$. 
    (b)
    Statistics of the correction term $c_{i,j}^{x_p}$ for each $x_p$.
    The red ($i=1$) and blue ($i=2$) boxes represent the 1st and 3rd quartiles of dataset $\{c_{i,j}^{x_p}\}_j$. 
    The horizontal lines mark the medians, while the dots mark the averages. 
    The whiskers extend to show the maximum and minimum values.
    (c)
    Histograms of $\phi_1-\phi_2$ for $x_p = 10$ (yellow), $20$ (green), $30$ (red), and $40$ (blue).
    The histograms are calculated over a duration in which $|\phi_1-\phi_2|$ increases by $200\pi$.
    The distribution calculated from the phase equations is displayed in gray.
    }
    \label{fig:fig6}
    \end{center}
\end{figure*}

\begin{figure}[h]
    \begin{center}
    \includegraphics[scale=1.0]{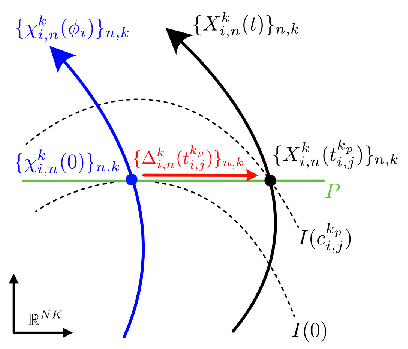}
    \caption{
    Illustration of two states on the same Poincar\'e section.
    We consider the isochron mapped from an infinite-dimensional space to a $KN$-dimensional space.  
    The two states, $\{\chi_{i,n}^k(0)\}_{n,k}$ and $\{X_{i,n}^k(t_{i,j}^{k_p})\}_{n,k}$, lie on the same Poincar\'e section $P$ but on different isochrons, $I(0)$ and $I(c_{i,j}^{k_p})$, respectively.
    The difference between these states is represented as $\{{\Delta}_{i,n}^k(t_{i,j}^{k_p})\}_{n,k}$ used in Eq.~(\ref{eq:PCA_correction_term}).
    These states are located on distinct isochrons unless the Poincar\'e section is the same as $I(0)$. 
    The correction term quantifies the difference between the phases assigned to these distinct isochrons. 
    }
    \label{fig:fig7}
    \end{center}
\end{figure}

\begin{figure*}[h]
    \begin{center}
    \includegraphics[scale=1.0]{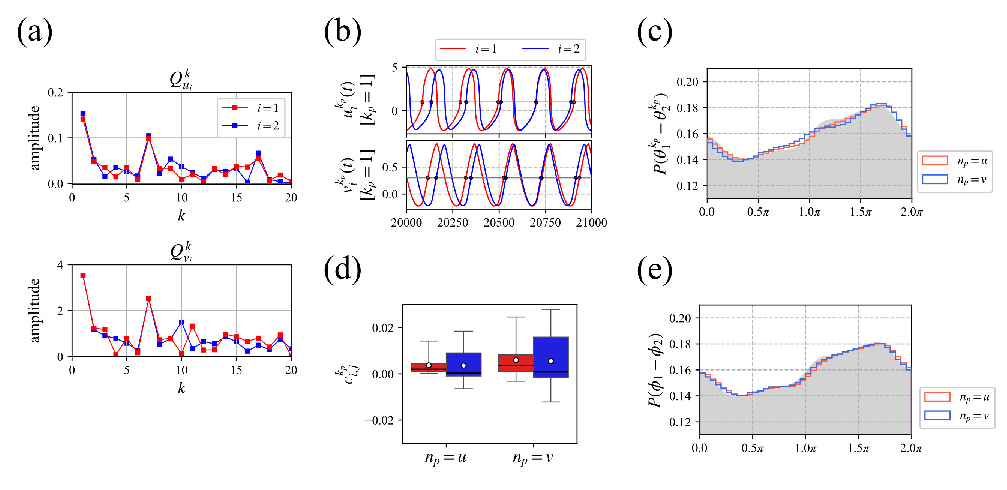}
    \caption{
    Calculation of the phase of spatiotemporal dynamics exhibiting target waves from measurements on all spatial grid points. 
    We adopt $k_p=1$.
    (a)
    Amplitude of $Q_{u_i}^k$ (top) and $Q_{v_i}^k$ (bottom).
    The Amplitude of $Q_{u_i}^k$ is calculated by $\int_0^{2\pi}|Q_{u_i}^k(\psi)| \mathrm{d} \psi /2\pi$ (a similar formula is used for $Q_{v_i}^k$).
    The red and blue lines indicate $i=1$ and $i=2$, respectively.
    (b)
    Time series of $u_i^{k_p}(t)$ (top) and $v_i^{k_p}(t)$ (bottom) and the Poincar\'e sections. 
    The red and blue lines indicate $i=1$ and $i=2$, respectively.
    The horizontal lines represent the Poincar\'e sections, and the dots represent the times when $u_i^{k_p}(t)$ or $v_i^{k_p}(t)$ intersects the Poincar\'e section from negative to positive.
    (c)
    Histograms of $\theta_1^{k_p}-\theta_2^{k_p}$ for $n_p=u$ (red) and $n_p=v$ (blue).
    The histograms are calculated over a duration in which $|\theta_1^{k_p}-\theta_2^{k_p}|$ increases by $200\pi$.
    The distribution calculated from the phase equations is displayed in gray. 
    (d)
    Statistics of the correction term $c_{i,j}^{k_p}$ for both $n_p = u$ and $n_p = v$.
    The red ($i=1$) and blue ($i=2$) boxes represent the 1st and 3rd quartiles of dataset $\{c_{i,j}^{k_p}\}_j$. 
    The horizontal lines mark the medians, while the dots mark the averages. 
    The whiskers extend to show the maximum and minimum values.
    (e)
    Histograms of $\phi_1-\phi_2$ for $n_p=u$ (red) and $n_p=v$ (blue).
    The histograms are calculated over a duration in which $|\phi_1-\phi_2|$ increases by $200\pi$.    
    The distribution calculated from the phase equations is displayed in gray.
    }
    \label{fig:fig8}
    \end{center}
\end{figure*}

\begin{figure*}[h]
    \begin{center}
    \includegraphics[scale=1.0]{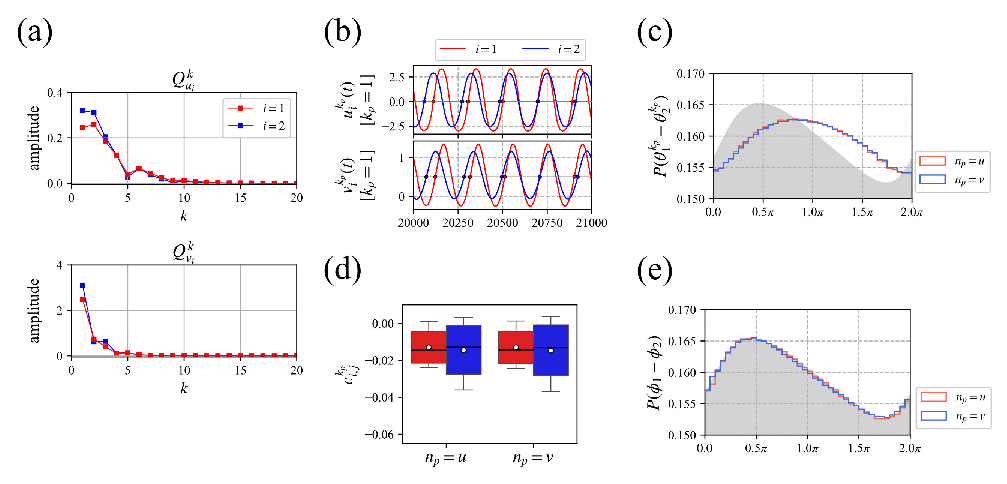}
    \caption{
    Calculation of the phase of spatiotemporal dynamics exhibiting oscillating spots from measurements on all spatial grid points. 
    We adopt $k_p=1$.
    (a)
    Amplitude of $Q_{u_i}^k$ (top) and $Q_{v_i}^k$ (bottom).
    The Amplitude of $Q_{u_i}^k$ is calculated by $\int_0^{2\pi}|Q_{u_i}^k(\psi)| \mathrm{d} \psi /2\pi$ (a similar formula is used for $Q_{v_i}^k$).
    The red and blue lines indicate $i=1$ and $i=2$, respectively.
    (b)
    Time series of $u_i^{k_p}(t)$ (top) and $v_i^{k_p}(t)$ (bottom) and the Poincar\'e sections. 
    The red and blue lines indicate $i=1$ and $i=2$, respectively.
    The horizontal lines represent the Poincar\'e sections, and the dots represent the times when $u_i^{k_p}(t)$ or $v_i^{k_p}(t)$ intersects the Poincar\'e section from negative to positive.
    (c)
    Histograms of $\theta_1^{k_p}-\theta_2^{k_p}$ for $n_p=u$ (red) and $n_p=v$ (blue).
    The histograms are calculated over a duration in which $|\theta_1^{k_p}-\theta_2^{k_p}|$ increases by $200\pi$.
    The distribution calculated from the phase equations is displayed in gray. 
    (d)
    Statistics of the correction term $c_{i,j}^{k_p}$ for both $n_p = u$ and $n_p = v$.
    The red ($i=1$) and blue ($i=2$) boxes represent the 1st and 3rd quartiles of dataset $\{c_{i,j}^{k_p}\}_j$. 
    The horizontal lines mark the medians, while the dots mark the averages. 
    The whiskers extend to show the maximum and minimum values.
    (e)
    Histograms of $\phi_1-\phi_2$ for $n_p=u$ (red) and $n_p=v$ (blue).
    The histograms are calculated over a duration in which $|\phi_1-\phi_2|$ increases by $200\pi$.    
    The distribution calculated from the phase equations is displayed in gray.
    }
    \label{fig:fig9}
    \end{center}
\end{figure*}

\begin{figure*}[h]
    \begin{center}
    \includegraphics[scale=1.0]{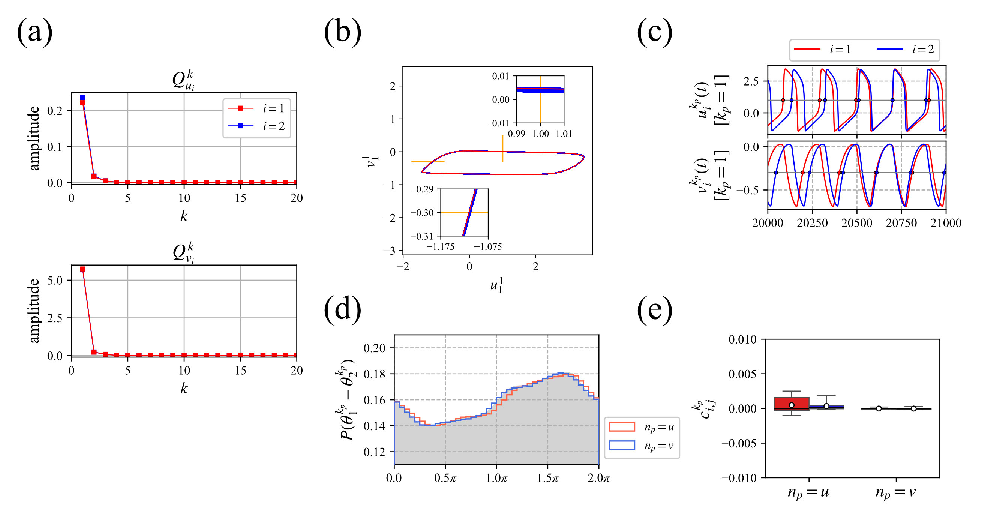}
    \caption{
    Calculation of the phase of spatiotemporal dynamics exhibiting target waves from measurements on all spatial grid points with the scheme described in Sec.~\ref{sec:PCA_QuQv}.
    We adopt $k_p=1$.
    (a)
    Amplitude of $Q_{u_i}^k$ (top) and $Q_{v_i}^k$ (bottom).
    The Amplitude of $Q_{u_i}^k$ is calculated by $\int_0^{2\pi}|Q_{u_i}^k(\psi)| \mathrm{d} \psi /2\pi$ (a similar formula is used for $Q_{v_i}^k$).
    The red and blue lines indicate $i=1$ and $i=2$, respectively.
    The amplitudes of $Q_{u_i}^k$ and $Q_{v_i}^k$ are localized to $k=1$.
    (b)
    Trajectory of $(u_1^1, v_1^1)$ (blue) and the limit-cycle of $(\chi_{u_1}^1, \chi_{v_1}^1)$ (red).
    The Poincar\'e sections for $n_p=u$ and $n_p=v$ are depicted with yellow lines, and the insets provide a close-up view around the intersection.
    (c) %%
    Time series of $u_i^{k_p}(t)$ (top) and $v_i^{k_p}(t)$ (bottom) and the Poincar\'e sections. 
    The red and blue lines represent $i=1$ and $i=2$, respectively.
    The horizontal lines represent the Poincar\'e sections, and the dots represent the times when $u_i^{k_p}(t)$ or $v_i^{k_p}(t)$ intersects the Poincar\'e section from negative to positive.
    (d)
    Histograms of $\theta_1^{k_p}-\theta_2^{k_p}$ for $n_p=u$ (red) and $n_p=v$ (blue).
    The histograms are calculated over a duration in which $|\theta_1^{k_p}-\theta_2^{k_p}|$ increases by $200\pi$.
    The distribution calculated from the phase equations is displayed in gray. 
    (e)
    Statistics of the correction term $c_{i,j}^{k_p}$ for both $n_p = u$ and $n_p = v$.
    The red ($i=1$) and blue ($i=2$) boxes represent the 1st and 3rd quartiles of dataset $\{c_{i,j}^{k_p}\}_j$. 
    The horizontal lines mark the medians, while the dots mark the averages. 
    The whiskers extend to show the maximum and minimum values.
    }
    \label{fig:fig10}
    \end{center}
\end{figure*}

\begin{figure*}[h]
    \begin{center}
    \includegraphics[scale=1.0]{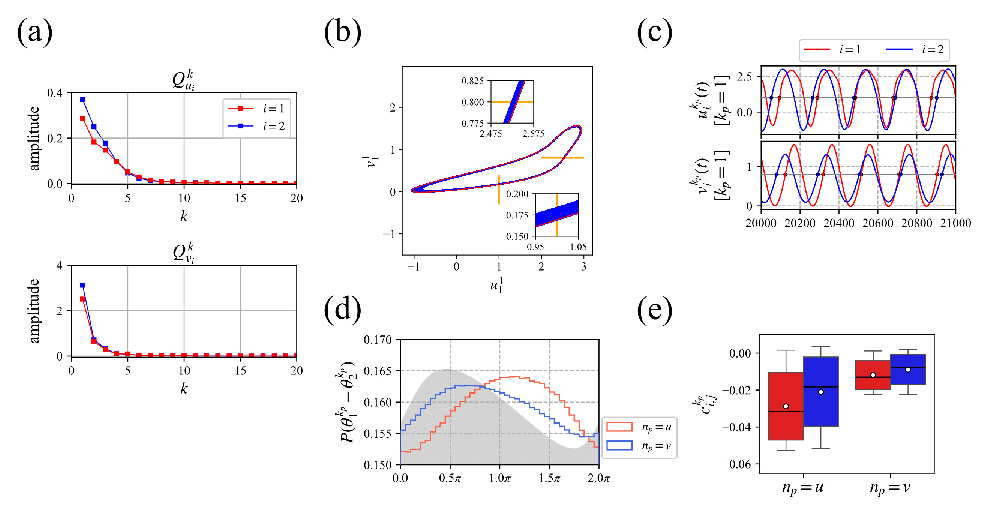}
    \caption{
    Calculation of the phase of spatiotemporal dynamics exhibiting oscillating spots from measurements on all spatial grid points with the scheme described in Sec.~\ref{sec:PCA_QuQv}.
    We adopt $k_p=1$.
    (a)
    Amplitude of $Q_{u_i}^k$ (top) and $Q_{v_i}^k$ (bottom).
    The Amplitude of $Q_{u_i}^k$ is calculated by $\int_0^{2\pi}|Q_{u_i}^k(\psi)| \mathrm{d} \psi / 2\pi$ (a similar formula is used for $Q_{v_i}^k$).
    The red and blue lines indicate $i=1$ and $i=2$, respectively.
    The amplitudes are not localized to a small number of $Q_{u_i}^k$ and $Q_{v_i}^k$ despite the implementation of the scheme described in Sec.~\ref{sec:PCA_QuQv}.
    (b)
    Trajectory of $(u_1^1, v_1^1)$ (blue) and the limit-cycle of $(\chi_{u_1}^1, \chi_{v_1}^1)$ (red).
    The Poincar\'e sections for $n_p=u$ and $n_p=v$ are depicted with yellow lines, and the insets provide a close-up view around the intersection.
    (c)
    Time series of $u_i^{k_p}(t)$ (top) and $v_i^{k_p}(t)$ (bottom) and the Poincar\'e sections. 
    The red and blue lines represent $i=1$ and $i=2$, respectively.
    The horizontal lines represent the Poincar\'e sections, and the dots represent the times when $u_i^{k_p}(t)$ or $v_i^{k_p}(t)$ intersects the Poincar\'e section from negative to positive.
    (d)
    Histograms of $\theta_1^{k_p}-\theta_2^{k_p}$ for $n_p=u$ (red) and $n_p=v$ (blue).
    The histograms are calculated over a duration in which $|\theta_1^{k_p}-\theta_2^{k_p}|$ increases by $200\pi$.
    The distribution calculated from the phase equations is displayed in gray. 
    (e)
    Statistics of the correction term $c_{i,j}^{k_p}$ for both $n_p = u$ and $n_p = v$.
    The red ($i=1$) and blue ($i=2$) boxes represent the 1st and 3rd quartiles of dataset $\{c_{i,j}^{k_p}\}_j$. 
    The horizontal lines mark the medians, while the dots mark the averages. 
    The whiskers extend to show the maximum and minimum values.
    }
    \label{fig:fig11}
    \end{center}
\end{figure*}

%%%%%%%%%%%%%%%%%%%%%%%%%%%%%%%%%%%%%%%%%%%%%%%%%%%%%%%%%%%%%
%%%%%%%%%%%%%%%%%%%%%%%%%%%%%%%%%%%%%%%%%%%%%%%%%%%%%%%%%%%%%
%%%%%%%%%%%%%%%%%%%%%%%%%%%%%%%%%%%%%%%%%%%%%%%%%%%%%%%%%%%%%
%%%%%%%%%%%%%%%%%%%%%%%%%%%%%%%%%%%%%%%%%%%%%%%%%%%%%%%%%%%%%
%%%%%%%%%%%%%%%%%%%%%%%%%%%%%%%%%%%%%%%%%%%%%%%%%%%%%%%%%%%%%

\begin{acknowledgments}
This work was supported by JSPS (Japan) KAKENHI Grant Numbers JP24K06910, JP20K03797, JP23H04467.
Numerical simulations were conducted using Earth Simulator at JAMSTEC.
\end{acknowledgments}

%%%%%%%%%%%%%%%%%%%%%%%%%%%%%%%%%%%%%%%%%%%%%%%%%%%%%%%%%%%%%
%%%%%%%%%%%%%%%%%%%%%%%%%%%%%%%%%%%%%%%%%%%%%%%%%%%%%%%%%%%%%
%%%%%%%%%%%%%%%%%%%%%%%%%%%%%%%%%%%%%%%%%%%%%%%%%%%%%%%%%%%%%
%%%%%%%%%%%%%%%%%%%%%%%%%%%%%%%%%%%%%%%%%%%%%%%%%%%%%%%%%%%%%
%%%%%%%%%%%%%%%%%%%%%%%%%%%%%%%%%%%%%%%%%%%%%%%%%%%%%%%%%%%%%

\appendix
\section{CALCULATION OF THE ISOCHRON-BASED PHASE BY LINEAR APPROXIMATION IN THE VICINITY OF LIMIT-CYCLE SOLUTION OF ODE}
\label{appendix:isochron_ODE}

\par 
Let us consider the following ODE:
\begin{align}
    \label{eq:ODEmodel}
    \dot{\bm{X}}(t) = \bm{F}(\bm{X}), 
\end{align}
where $\bm{X} \in \mathbb{R}^N$ represents the state variable. 
We assume that Eq.(\ref{eq:ODEmodel}) has a limit-cycle solution $\bm{\chi}$ with period $T$.
According to phase reduction theory \cite{Winfree_1980, Kuramoto_1984}, the multidimensional state space is mapped to a one-dimensional space characterized by a periodic variable, $\phi \in [0, 2\pi)$, called phase.
Here, let us consider a state, $\bm{X}_0(t)$, which evolves along the limit-cycle solution.  
The phase $\phi(t)$ is assigned to $\bm{X}_0(t)$ such that it increases linearly with a constant frequency $\omega:=2\pi/T$ as time progresses, i.e., $\phi(t)=\omega t$.
We define the state on the limit-cycle solution as $\bm{\chi}(\phi(t)) := \bm{X}_0(t)$. 
Next, we extend the definition of the phase to the basin of attraction.
We assign the same phase value to a subset of the state space defined as follows:
\begin{align}
    \label{eq:ODEconvergence}
    & I(\psi)  
    = \bigg\{\bm{X}(t)  \bigg| \lim_{t \to +\infty} \| \bm{X}(t) - \bm{\chi}(\phi(t)) \| = 0, ~
    \phi(0)=\psi  \bigg\},
\end{align}
where $\| \cdot \|$ denotes the $L^2$ norm defined as $\|\bm{A} \| = \sqrt{\bm{A} \cdot \bm{A}}$. 
The subset assigned phase value $\phi$ is represented as $I(\phi)$.
By analyzing the isochrons corresponding to each phase value, we can determine a scalar field that represents the configuration of the phase over the basin of attraction.
The time series of the phase $\phi(t)$ is obtained by assigning the phase value to the time series of $\bm{X}(t)$ on the basis of this scalar field.

\par
The concept of the isochron provides a linear approximation in the vicinity of the limit-cycle solution.
We consider a state $\bm{X}'$ which is slightly kicked out from the state $\bm{\chi}(\psi_0)$ by a weak perturbation. 
The phase, $\psi$, assigned to $\bm{X}'$ is calculated as follows:
\begin{align}
    \label{eq:ODEperturbation}
    \psi 
    & = \psi_0 + \bm{Z}(\psi_0) \cdot \left(\bm{X}' - \bm{\chi}(\psi_0) \right)   \nonumber \\
    & = \psi_0 + \sum_{n=1}^N Z_n(\psi_0) \left( X_n'- {\chi}_n(\psi_0) \right),
\end{align}
where $\bm{Z}(\psi_0) \in \mathbb{R}^N$ is the phase sensitivity function, which quantifies linear response characteristics of the phase to weak perturbation, and the subscript $n$ denotes the $n$th entry of the vector.
When the two states, $\bm{X}'$ and $\bm{\chi}(\psi_0)$, lie on the same Poincar\'e section that defines $\psi_0 = 0$, the second term can be considered as a correction term, and $\psi$ is interpreted as the isochron-based phase.

\section{CALCULATION OF THE ISOCHRON-BASED PHASE BY LINEAR APPROXIMATION IN THE VICINITY OF LIMIT-CYCLE SOLUTION OF PDE}
\label{appendix:isochron}

\par 
Similar to the ODE described in Appendix~\ref{appendix:isochron_ODE}, the concept of isochron can be applied to the PDE~\cite{Nakao_2014}.
Let us consider the equation described in Eq.~(\ref{eq:RDmodel}) without the coupling ($\bm{G}=\bm{0}$) as follows (the subscript $i$ is removed for convenience):
\begin{align}
    \label{eq:RDmodel_WithoutCoupling}
    \frac{\partial}{\partial t} {\bm{X}}(\bm{r},t) = \bm{F}(\bm{X}, \bm{r}) + D \nabla^2 \bm{X}. 
\end{align}
We assume that Eq.(\ref{eq:RDmodel_WithoutCoupling}) has a limit-cycle solution $\bm{\chi}$ with period $T$.
According to the phase reduction theory extended to the PDE~\cite{Nakao_2014}, the infinite-dimensional state space is mapped to a one-dimensional space characterized by a periodic variable, $\phi \in [0, 2\pi)$.
Here, let us consider a state, $\bm{X}_0(\bm{r}, t)$, which evolves along the limit-cycle solution.  
The phase $\phi(t)$ is assigned on $\bm{X}_0(\bm{r}, t)$ such that it increases linearly with constant frequency $\omega:=2\pi/T$ as time progresses, i.e., $\phi(t)=\omega t$.
We define a state on the limit-cycle solution as $\bm{\chi}(\bm{r}, \phi(t)) := \bm{X}_0(\bm{r},t)$. 
Next, we extend the definition of the phase to the basin of attraction.
We assign the same phase value to a subset of the state space as follows:
\begin{align}
    \label{eq:convergence}
    & I(\psi) = \nonumber \\
    & \bigg\{\bm{X}(\bm{r},t) \bigg| \lim_{t \to +\infty} \| \bm{X}(\bm{r},t) - \bm{\chi}(\bm{r},\phi(t)) \| = 0, 
    ~ \phi(0)=\psi  \bigg\},
\end{align}
where $\| \cdot \|$ denotes the $L^2$ norm defined as $\| \bm{A}(\bm{r}) \| = \sqrt{\int \bm{A}(\bm{r}) \cdot \bm{A}(\bm{r}) \mathrm{d}\bm{r} }$. 
We denote the subset assigned phase value $\phi$ as $I(\phi)$.
By analyzing the isochrons corresponding to each phase value, we can determine a scalar field that represents the configuration of the phase over the basin of attraction.
Therefore, the time series of the phase $\phi(t)$ is obtained by assigning the phase value to the time series of $\bm{X}(\bm{r},t)$ on the basis of this scalar field.

\par
The concept of the isochron provides a linear approximation in the vicinity of the limit-cycle solution.
We consider a state $\bm{X}'(\bm{r})$ which is slightly kicked out from a state $\bm{\chi}(\bm{r}, \psi_0)$ by a weak perturbation. 
The phase, $\psi$, assigned to $\bm{X}'$ is calculated as follows:
\begin{align}
    \label{eq:perturbation}
    \psi 
    & = \psi_0 + \int \bm{Q}(\bm{r},\psi_0) \cdot \left(\bm{X}'(\bm{r}) - \bm{\chi}(\bm{r}, \psi_0) \right) \mathrm{d}\bm{r} \nonumber \\
    & = \psi_0 + \int \sum_{n=1}^N Q_n(\bm{r}, \psi_0) \left( X_n'(\bm{r}) - {\chi}_n(\bm{r}, \psi_0) \right) \mathrm{d}\bm{r},
\end{align}
where $\bm{Q}$ is the phase sensitivity function and the subscript $n$ denotes the $n$th entry of the vector.
For simplicity, we assume a one-dimensional space, i.e., $\bm{r} \to x$.
Given the spatial grid, $x_m = m\Delta x~(m=0,1,\ldots,M)$, Eq.~(\ref{eq:perturbation}) is rewritten with discrete representation as follows:
\begin{align}
    \label{eq:perturbation2}
    \psi 
    & = \psi_0 + \sum_{m=0}^M \sum_{n=1}^N Q_n(x_m, \psi_0) \left( X_n'(x_m) - {\chi}_n(x_m, \psi_0) \right) \Delta x_m , \\
    & \Delta x_m := 
    \begin{cases}
        \Delta x / 2 & (m=0,M), \\
        \Delta x     & (\mathrm{otherwise}). \\
    \end{cases} \nonumber
\end{align}
The calculation of the correction term in Eq.~(\ref{eq:1d_correction_term}) is derived from this equation.
When we replace the variables in Eq.~(\ref{eq:perturbation2}) with 
$\psi \to 2\pi j + c_{i,j}^{x_p}$, 
$\psi_0 \to 2\pi j$,
$Q_n(x_m, \psi_0) \to Q_{i,n}(x_m, 0)$,
$X_n'(x_m) \to X_{i,n}(x_m, t_{i,j}^{x_p})$, 
and $\chi_n(x_m, \psi_0) \to \chi_{i,n}(x_m, 0)$ and then consider $X_{i,n_p}(x_{m_p}, t_{i,j}^{x_p}) - \chi_{i,n_p}(x_{m_p}, 0) = 0$, we obtain Eq.~(\ref{eq:1d_correction_term}).

\section{LIMIT-CYCLE SOLUTION AND PHASE SENSITIVITY FUNCTION OF THE FHN REACTION-DIFFUSION MODEL}
\label{appendix:limitcycle}

\par
\begin{figure*}[h]
    \begin{center}
        \includegraphics[scale=1.0]{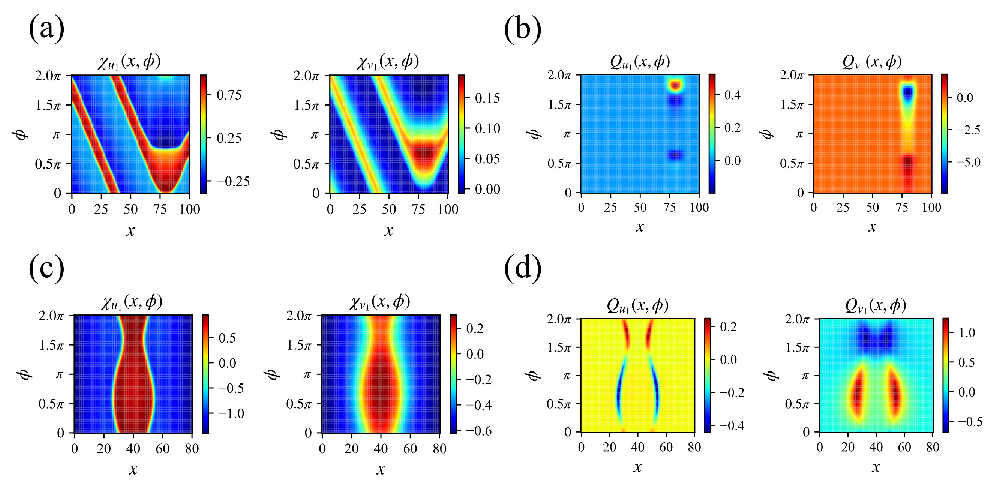}
        \caption{
        Limit-cycle solutions and phase sensitivity functions for the FHN reaction-diffusion model.
        (a)
        Limit cycle of the target-wave solution  
        $\bm{\chi}_1(x, \phi)=\left( \chi_{u_1}(x,\phi), \chi_{v_1}(x, \phi) \right)$.
        (b)
        Phase sensitivity function of the target-wave solution 
        $\bm{Q}_1(x, \phi)=\left( Q_{u_1}(x,\phi), Q_{v_1}(x, \phi) \right)$.
        (c)
        Limit cycle of the oscillating spot solution $\bm{\chi}_1(x, \phi)$.
        (d)
        Phase sensitivity function of the oscillating spot solution $\bm{Q}_1(x, \phi)$}.
        \label{fig:figA1}
    \end{center}
\end{figure*}

 
The FHN reaction-diffusion model described in Secs.~\ref{sec:model_TW} and \ref{sec:model_OS} has the limit-cycle solution $\bm{\chi}_i(x, \phi_i) = \left(\chi_{u_i}(x, \phi_i), \chi_{v_i}(x, \phi_i) \right)$ and the phase sensitivity function $\bm{Q}_i(x, \phi_i)=\left(Q_{u_i}(x, \phi_i), Q_{v_i}(x, \phi_i) \right)$. 
The limit-cycle solution $\bm{\chi}_1$ and the phase sensitivity function $\bm{Q}_1$ for the target-wave solution are shown in Figs.~\ref{fig:figA1}(a) and \ref{fig:figA1}(b), and those for the oscillating spot solution are shown in Figs.~\ref{fig:figA1}(c) and \ref{fig:figA1}(d).

\section{PHASE EQUATIONS OF THE FHN REACTION-DIFFUSION MODEL}
\label{appendix:phase_equation}

\par 
Phase reduction theory extended to the PDE~\cite{Nakao_2014} allows for deriving a phase equation from a PDE with a limit-cycle solution. 
The FHN reaction-diffusion model falls within the applicability of this theory. 
For a pair of weakly coupled reaction-diffusion models described by Eq.~(\ref{eq:RDmodel}), the phase equations are given by
\begin{align}
    \label{eq:phase_equation}
    \begin{split}
    \dot{\phi}_1(t) = \omega_1 + \Gamma_{12}(\phi_1-\phi_2), \\
    \dot{\phi}_2(t) = \omega_2 + \Gamma_{21}(\phi_2-\phi_1),
    \end{split}
\end{align}
where $\omega_i:= 2\pi / T_i$ represents the frequency of the limit-cycle solution $\bm{\chi}_i$ of Eq.~(\ref{eq:RDmodel}) without coupling ($\bm{G}=\bm{0}$). 
The constant $T_i$ represents the period of the limit-cycle solution.
The phase coupling functions between the two phases, $\Gamma_{12}$ and $\Gamma_{21}$, describe how the phases are affected by the coupling function $\bm{G}$.
The phase equations for the coupled FHN reaction-diffusion models are displayed in Fig.~\ref{fig:figA2} in Appendix~\ref{appendix:estimation}.
(Figs.~\ref{fig:figA3} and \ref{fig:figA4} in Appendix~\ref{appendix:estimation} also show the same.)

\par 
The distribution of $\phi_1-\phi_2$ shown in several figures is calculated using the phase equations (see Figs.~\ref{fig:fig2}(c), \ref{fig:fig4}(c), \ref{fig:fig5}(b), \ref{fig:fig6}(c), \ref{fig:fig8}(c), \ref{fig:fig8}(e), \ref{fig:fig9}(c), \ref{fig:fig9}(e), \ref{fig:fig10}(d), and \ref{fig:fig11}(d)).
The distribution is inversely related to the velocity of $\phi_1-\phi_2$. 
Therefore, we calculated it by taking the reciprocal of the difference between the phase equations, i.e., $P(\phi_1-\phi_2) \propto 1 /\left( \omega_1 - \omega_2 + \Gamma_{12}(\phi_1-\phi_2) - \Gamma_{21}(-(\phi_1-\phi_2)) \right)$.

\section{ESTIMATING THE PHASE EQUATIONS}
\label{appendix:estimation}

\par 
Ignoring the correction term when calculating the phase causes errors in estimating the phase equations (Eq.~(\ref{eq:phase_equation})).
This error becomes more significant when the magnitude of the correction term is large.
For illustration, we estimated the phase equation using the method developed in Ref.~\cite{Ota_2014}.

\par
\begin{figure*}[h]
    \begin{center}
        \includegraphics[scale=1.0]{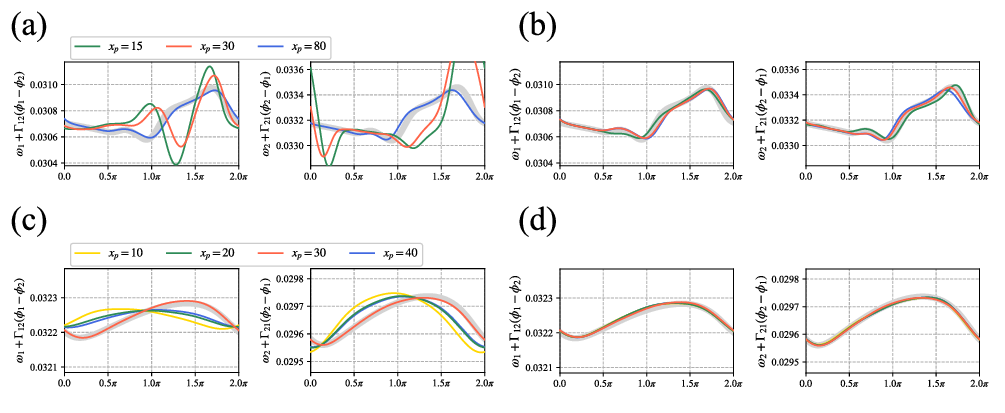}
        \caption{
        Phase equations estimated from the time series of the phase calculated from measurements taken at a single spatial grid point (see Secs.~\ref{sec:1d_TW} and \ref{sec:1d_OS}).
        The colored lines display the phase equations estimated from the time series of $\theta_i^{x_p}(t)$ or $\phi_i(t)$. 
        The legend denotes the correspondence between the colors of the lines and the values of $x_p$.
        The true forms of the phase equations are displayed in bold gray lines.
        (a)
        Results for target waves obtained from $\theta_i^{x_p}$.
        (b)
        Results for target waves obtained from $\phi_i$.
        (c)
        Results for oscillating spots obtained from $\theta_i^{x_p}$.
        (d)
        Results for oscillating spots obtained from $\phi_i$.
        }
        \label{fig:figA2}
    \end{center}
\end{figure*}

 
Figures~\ref{fig:figA2}(a) and \ref{fig:figA2}(b) show the phase equations estimated from the time series of $\theta_i^{x_p}(t)$ and $\phi_i(t)$, respectively, for spatiotemporal dynamics exhibiting target waves.
Figures~\ref{fig:figA2}(c) and \ref{fig:figA2}(d) show the same estimations for dynamics exhibiting oscillating spots.
We examine the results for $x_p \simeq 15, 30$ shown in Fig.~\ref{fig:figA2}(a) and those for $x_p=10, 20, 40$ shown in Fig.~\ref{fig:figA2}(c). 
In these cases, the magnitude of the correction term is substantial (see Figs.~\ref{fig:fig4}(b) and \ref{fig:fig6}(b)), leading to a significant discrepancy between the estimated phase equations and the true forms indicated by bold gray lines.
Next, we focus on the results for $x_p \simeq 80$ shown in Fig.~\ref{fig:figA2}(a) and those for $x_p=30$ shown in Fig.~\ref{fig:figA2}(c).
The magnitude of the correction term is small in these cases, and the estimated phase equations closely match the true forms.
In summary, these results indicate that ignoring the correction term impacts the estimation results, and the extent of the impact depends on the magnitude of the correction term.
Furthermore, the phase equations estimated from the time series of $\phi_i(t)$ also closely resemble the true forms (Figs.~\ref{fig:figA2}(b) and \ref{fig:figA2}(d)).
Therefore, incorporating correction terms in phase calculation effectively reduces the estimation errors in the phase equations.

\par
\begin{figure*}[h]
    \begin{center}
        \includegraphics[scale=1.0]{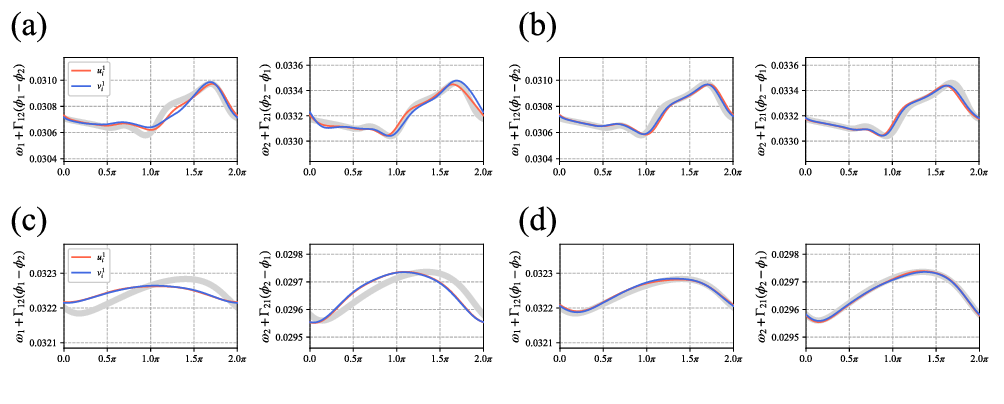}
        \caption{
        Phase equations estimated from the time series of the phase calculated from measurements taken at all spatial grid points. (see Sec.~\ref{sec:PCA_uv}).
        The red and blue lines display the phase equations obtained from the time series of $u_i^{k_p}(t)$ and $v_i^{k_p}(t)$ ($k_p=1$), respectively.
        The true forms of the phase equations are displayed in bold gray lines.
        (a)
        Results for target waves obtained from $\theta_i^{k_p}$.
        (b)
        Results for target waves obtained from $\phi_i$.
        (c)
        Results for oscillating spots obtained from $\theta_i^{k_p}$.
        (d)
        Results for oscillating spots obtained from $\phi_i$.
        }
        \label{fig:figA3}
    \end{center}
\end{figure*}

The same equations are estimated in Sec.~\ref{sec:PCA_uv}.
Figures~\ref{fig:figA3}(a) and \ref{fig:figA3}(b) show the phase equations estimated from the time series of $\theta_i^{k_p}(t)$ and $\phi_i(t)$, respectively, for spatiotemporal dynamics exhibiting target waves. 
Figures~\ref{fig:figA3}(c) and \ref{fig:figA3}(d) show the same estimations for spatiotemporal dynamics exhibiting oscillating spots.
We adopt $k_p=1$ as we do in Secs.~\ref{sec:PCA_uv} and Sec.~\ref{sec:PCA_QuQv}. 
According to Figs.~\ref{fig:fig8}(d) and \ref{fig:fig9}(d), the magnitude of the correction term is small for target waves but large for oscillating spots.
The results shown in Figs.~\ref{fig:figA3}(a) closely resemble the true forms, although those shown in Figs.~\ref{fig:figA3}(c) significantly differ from the true forms.
These observations reaffirm that larger correction terms lead to greater errors in phase equation estimation.
Furthermore, the phase equations estimated from the time series of $\phi_i(t)$ shown in Figs.~\ref{fig:figA3}(b) and \ref{fig:figA3}(d) are more accurate compared to those estimated from the time series of $\theta_i^{k_p}(t)$.

\par
\begin{figure}[h]
    \begin{center}
        \includegraphics[scale=1.0]{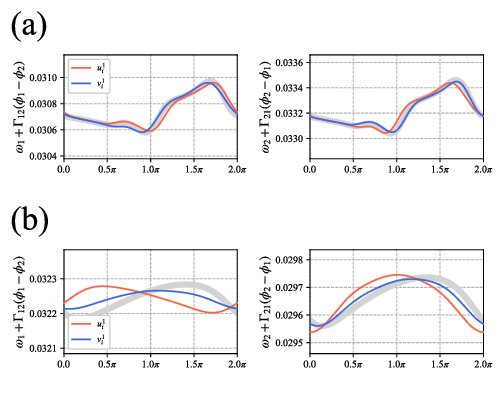}
        \caption{
        Phase equations estimated from the time series of the phase using the scheme described in Sec.~\ref{sec:PCA_QuQv}.
        The red and blue lines display the phase equations obtained from the time series of $u_i^{k_p}(t)$ and $v_i^{k_p}(t)$ ($k_p=1$), respectively.
        The true forms of the phase equations are displayed in bold gray lines.
        (a)
        Results for target waves obtained from $\theta_i^{k_p}$.
        (b)
        Results for oscillating spots obtained from $\theta_i^{k_p}$.
        }
        \label{fig:figA4}
    \end{center}
\end{figure}

 
The same equations are estimated again in Sec.~\ref{sec:PCA_QuQv}.
Figures~\ref{fig:figA4}(a) and \ref{fig:figA4}(b) show the phase equations estimated from the time series of $\theta_i^{k_p}(t)$ for spatiotemporal dynamics exhibiting target waves and oscillating spots, respectively.
Figures~\ref{fig:figA3}(a) and \ref{fig:figA4}(a) both show phase equations estimated from $\theta_i^{k_p}(t)$ for target waves but utilize different decomposition schemes (see Secs.~\ref{sec:PCA_uv} and \ref{sec:PCA_QuQv}). 
The latter estimation appears to capture more detailed features of the phase equations compared to the former estimation.
Similarly, Figs.~\ref{fig:figA3}(c) and \ref{fig:figA4}(b) present phase equations estimated from $\theta_i^{k_p}(t)$ for oscillating spots but utilize different decomposition schemes.
In both schemes, the estimations show significant deviations from the true forms. 
These results indicate that the scheme described in Sec.~\ref{sec:PCA_QuQv} reduced the magnitude of the correction term and improved the estimation of the phase equations only for the case of target waves.

\par 
The Detail of the method of the estimation is left to Ref.~\cite{Ota_2014}. 
The hyperparameters are set to  
$\chi_i^{\mathrm{old}}=\{\bar{\omega}_i,0,0,\ldots,0 \}$, 
$\Sigma_i^{\mathrm{old}}=\lambda_i^{-1} E$, 
and $\alpha_i^{\mathrm{old}}= \beta_i^{\mathrm{old}} = 0.001$ (see Ref.~\cite{Ota_2014}), where $\bar{\omega}_i$ is the mean velocity calculated from the time series of the phase, and $E$ denotes an identity matrix.
In addition to the hyperparameters, we set two parameters, $\lambda_i$, which represents the magnitude of $\Sigma_i^{\mathrm{old}}$, and $M_i$, which represents the maximum order of the Fourier series used in the phase coupling function.
These parameters were optimized through model selection to maximize the marginal likelihood functions within the following ranges: $M_i=0,1,\ldots,10$ and $\log_{10}\lambda_i=1,2,\ldots,5$.
The phase equations were estimated from the time series of the phase during periods when $|\phi_1-\phi_2|$ increases by $200\pi$.

\section{THE ORTHOGONALLY DECOMPOSED PHASE SENSITIVITY FUNCTION}
\label{appendix:proof}

\par 
We mention the derivation of Eq.~(\ref{eq:PCA_correction_term}).
Let us consider the state variable $\bm{X}(x,t) \in \mathbb{R}^N$ which obeys 
\begin{align}
    \label{eq:Proof_GeneralForm}
    \frac{\partial }{\partial t} \bm{X}(x,t) 
    = \bm{F}(\bm{X}, x) + \epsilon \bm{p}(x,t),
\end{align}
where $\bm{F}(\bm{X}, x)$ represents the local dynamics at point $x$ and time $t$, $\bm{p}(x, t)$ represents the local perturbation to $\bm{X}(x,t)$, and $\epsilon \ll 1$ represents the intensity of the perturbation.
For simplicity, the local dynamics $\bm{F}$ includes any diffusion of $\bm{X}$ if present.
We assume that Eq.~(\ref{eq:Proof_GeneralForm}), in the absence of coupling ($\bm{p}=\bm{0}$), has a limit-cycle solution $\bm{\chi}(x, \phi)$ with $\phi=[0, 2\pi)$. 
Additionally, we assume that the perturbation is sufficiently weak so that $\bm{X}$ does not deviate significantly from the limit-cycle solution.
Given the phase sensitivity function $\bm{Q}(x, \phi)$ for the limit-cycle solution, the phase equation is derived as follows~\cite{Nakao_2014}:
\begin{align}
    \label{eq:Proof_Phasedynamics}
    \dot{\phi}(t) 
    & = \omega + \epsilon \int_0^L \bm{Q}(x, \phi) \cdot \bm{p}(x, t) \mathrm{d}x \nonumber \\
    & = \omega + \epsilon \int_0^L \sum_{n=1}^N Q_n(x, \phi) p_n(x, t) \mathrm{d}x,
\end{align}
where, subscript $n$ denotes the $n$th entry of the vectors, $L$ is the size of the system, and $\omega$ is frequency of the limit-cycle solution.

\par 
We consider mapping from an infinite-dimensional state space to the $KN$-dimensional space through orthogonal decomposition. 
Given a set of orthonormal basis functions $\Phi_n^k~(n=1,2,\ldots,N,~k=1,2,\ldots,K)$, the $N$-dimensional state or functions, 
$\tilde{\bm{X}}^k(t) = ( \tilde{X}_1^k(t), \tilde{X}_2^k(t), \ldots, \tilde{X}_N^k(t) )$, 
$\tilde{\bm{Q}}^k(\phi) = ( \tilde{Q}_1^k(\phi), \tilde{Q}_2^k(\phi), \ldots, \tilde{Q}_N^k(\phi) )$,
$\tilde{\bm{F}}^k(\bm{X}) = ( \tilde{F}_1^k(\bm{X}), \tilde{F}_2^k(\bm{X}), \ldots, \tilde{F}_N^k(\bm{X}) )$,
and $\tilde{\bm{p}}^k(t) = ( \tilde{p}_1^k(t), \tilde{p}_2^k(t), \ldots, \tilde{p}_N^k(t) )$ are obtained as follows:
\begin{align}
    \label{eq:Proof_projection}
    \begin{split}
    & \tilde{X}_n^k(t) = \int_0^L  X_n(x,t) \Phi_n^k(x) \mathrm{d}x, \\
    & \tilde{Q}_n^k(\phi) = \int_0^L  Q_n(x,\phi) \Phi_n^k(x) \mathrm{d}x, \\
    & \tilde{F}_n^k(\bm{X}) = \int_0^L F_n(\bm{X}, x) \Phi^k_n(x) \mathrm{d}x, \\
    & \tilde{p}_n^k(t) = \int_0^L  p_n(x,t) \Phi_n^k(x) \mathrm{d}x,
    \end{split}
\end{align}
where the basis function satisfies $\int_0^L \Phi^p_n(x) \Phi^q_n(x) = \delta_{pq}$.
The tilde indicates that the value is obtained by projection onto the basis functions.
We assume that the number of the component $K$ is sufficiently large to ensure that each function can be reproduced nearly $100\%$.
The dynamics of the state variable $X_n^k$ projected onto $\Phi_n^k(x)$ obeys
\begin{align}
    \label{eq:Proof_projected_dynamics}
    \frac{\partial }{\partial t} \tilde{X}_n^k(t)
    = \tilde{F}_n^k(\bm{X}) + \epsilon \tilde{p}_n^k(t).
\end{align}
We rewrite Eq.~(\ref{eq:Proof_projected_dynamics}) with $N$-dimensional vector representation as follows:
\begin{align}
    \label{eq:Proof_projected_dynamics2}
    \frac{\partial }{\partial t} \tilde{\bm{X}}^k(t) 
    = \tilde{\bm{F}}^k(\bm{X}) + \epsilon \bm{\tilde{p}}^k(t). 
\end{align}
Given Eq.~(\ref{eq:Proof_projected_dynamics2}) for $k=1,2,\ldots,K$, the state variable $\tilde{\bm{X}}(t) = (\tilde{\bm{X}}^1(t), \tilde{\bm{X}}^2(t), \ldots, \tilde{\bm{X}}^K(t) ) \in \mathbb{R}^{KN}$ is subjected to the perturbation $\tilde{\bm{p}}(t) = (\tilde{\bm{p}}^1(t), \tilde{\bm{p}}^2(t), \ldots, \tilde{\bm{p}}^K(t) ) \in \mathbb{R}^{KN}$.
The limit-cycle solution of Eq.~(\ref{eq:Proof_GeneralForm}) persists as the limit-cycle solution of Eq.~(\ref{eq:Proof_projected_dynamics2}), albeit it deformed. 
Therefore, there exists a phase equation that describes the phase response to the perturbation $\tilde{\bm{p}}$, as follows:
\begin{align}
    \label{eq:Proof_Phasedynamics_projected}
    \dot{\phi}(t)
    & = \omega + \epsilon \hat{\bm{Q}}(\phi) \cdot \tilde{\bm{p}}(t) \nonumber \\
    & = \omega + \epsilon \sum_{k=1}^K \hat{\bm{Q}}^k(\phi) \cdot \tilde{\bm{p}}^k(t) \nonumber \\
    & = \omega + \epsilon \sum_{k=1}^{K} \sum_{n=1}^N \hat{Q}_n^k(\phi) \tilde{p}_n^k(t), 
\end{align}
where unknown $\hat{\bm{Q}}(\phi) = ( \hat{\bm{Q}}^1(\phi), \hat{\bm{Q}}^2(\phi), \ldots, \hat{\bm{Q}}^K(\phi) ) \in \mathbb{R}^{KN}$ with $\hat{\bm{Q}}^k(\phi)= ( \hat{Q}_1^k(\phi), \hat{Q}_2^k(\phi), \ldots, \hat{Q}_N^k(\phi) )$ represents the linear response characteristics of the phase to the perturbation $\tilde{\bm{p}}$. 
The frequency $\omega$ in Eq.~(\ref{eq:Proof_Phasedynamics_projected}) is the same as that in Eq.~(\ref{eq:Proof_Phasedynamics}) since the period of the limit-cycle solution of Eq.~(\ref{eq:Proof_GeneralForm}) and Eq.~(\ref{eq:Proof_projected_dynamics2}) must have same period.

\par 
Here, we derive what the unknown function $\hat{\bm{Q}}$ is.
From Eqs.~(\ref{eq:Proof_Phasedynamics}) and (\ref{eq:Proof_Phasedynamics_projected}), we obtain the following equation:
\begin{align}
    \label{eq:Proof_A}
    \int_0^L
    \sum_{n=1}^N Q_n(x,\phi) p_n(x,t) \mathrm{d}x
    = \sum_{k=1}^K \sum_{n=1}^N \hat{Q}_n^k(\phi) \tilde{p}_n^k(t).
\end{align}
We also obtain the following equation starting from the left-hand side of Eq.~(\ref{eq:Proof_A}): 
\begin{align}
    \label{eq:Proof_B}
    & \int_0^L \sum_{n=1}^N Q_n(x,\phi) p_n(x,t) \mathrm{d}x \nonumber \\
    & = \int_0^L  \sum_{n=1}^N Q_n(x,\phi) \left[ \sum_{k=1}^K \tilde{p}_n^k(t) \Phi_n^k(x) \right] \mathrm{d}x \nonumber \\
    & = \sum_{k=1}^K \sum_{n=1}^N \left[ \int_0^L  Q_n(x,\phi) \Phi_n^k(x) \mathrm{d}x \right] \tilde{p}_n^k(t) \nonumber \\
    & = \sum_{k=1}^K \sum_{n=1}^N \tilde{Q}_n^k(\phi) \tilde{p}_n^k(t).
\end{align}
The transformation to the second and fourth rows is achieved by substituting $p_n(x,t) \simeq \sum_{k=1}^K \tilde{p}_n^k(t) \Phi_n^k(x)$ and $\tilde{Q}_n^k(t) = \int_0^L  Q_n(x,t) \Phi_n^k(x) \mathrm{d}x$, respectively.
Finally, we derived $\hat{Q}_n^k(\phi) = \tilde{Q}_n^k(\phi)$ from Eqs.~(\ref{eq:Proof_A}) and (\ref{eq:Proof_B}), and thus we obtain the following equation from Eq.~(\ref{eq:Proof_Phasedynamics_projected}):
\begin{align}
    \label{eq:Proof_DerivationResult}
    \dot{\phi}(t)
    & = \omega + \epsilon \sum_{k=1}^{K} \sum_{n=1}^N \tilde{Q}_n^k(\phi) \tilde{p}_n^k(t).
\end{align}
We also obtain the equation with vector representation as follows:
\begin{align}
    \label{eq:Proof_DerivationResult2}
    \dot{\phi}(t)
    & = \omega + \epsilon \sum_{k=1}^{K} \tilde{\bm{Q}}^k(\phi) \cdot \tilde{\bm{p}}^k(t) \nonumber \\
    & = \omega + \epsilon \tilde{\bm{Q}}(\phi) \cdot \tilde{\bm{p}}(t).
\end{align}
Equations~(\ref{eq:Proof_DerivationResult}) and (\ref{eq:Proof_DerivationResult2}) indicate that $\tilde{\bm{Q}}$ serves as the phase sensitivity function when the infinite-dimensional state space is mapped to the finite-dimensional space spanned by basis functions as described in Eq.~(\ref{eq:Proof_projection}).

\par
We denote the limit-cycle solution projected onto the basis functions as $\tilde{\bm{\chi}}(\phi) \in \mathbb{R}^{KN}$,  whose $n$th entry is calculated by $\tilde{\chi}_n^k(t) = \int_0^L  \chi_n(x,t) \Phi_n^k(x) \mathrm{d}x$.
In the previous paragraph, we found that $\tilde{\bm{Q}}$ represents the linear response characteristics of the phase to the perturbation $\tilde{\bm{p}}$ when $\tilde{\bm{Q}}$ and $\tilde{\bm{p}}$ are obtained by the projection onto the same basis functions.
Therefore, the phase $\psi$ assigned to a state $\tilde{\bm{X}}' \in \mathbb{R}^{KN}$, which is slightly kicked out from the state $\tilde{\bm{\chi}}(\psi_0)$, is calculated as follows:
\begin{align}
    \label{eq:Proof_perturbation}
    \psi 
    & = \psi_0 + \tilde{\bm{Q}}(\psi_0) \cdot \left(\tilde{\bm{X}}' - \tilde{\bm{\chi}}(\psi_0) \right)   \nonumber \\
    & = \psi_0 + \sum_{k=1}^K \sum_{n=1}^N \tilde{Q}_n^k(\psi_0) \left( [\tilde{X}']_n^k- {\chi}_n^k(\psi_0) \right).
\end{align}
The calculation of the correction term in Eq.~(\ref{eq:PCA_correction_term}) is based on this equation.
Specifically, when we replace the variables in Eq.~(\ref{eq:Proof_perturbation}) with 
$\psi \to 2\pi j + c_{i,j}^{k_p}$, 
$\psi_0 \to 2\pi j$,
$\tilde{Q}_n^k(\psi_0) \to Q_{i,n}^k(0)$,
$[\tilde{X}']_n^k \to X_{i,n}^k(t_{i,j}^{k_p})$, 
and $\tilde{\chi}_n^k(\psi_0) \to \chi_{i,n}^k(0)$ and then consider $X_{i,n_p}^{k_p}(t_{i,j}^{k_p}) - \chi_{i,n_p}^{k_p}(0) = 0$, we obtain Eq.~(\ref{eq:PCA_correction_term}).
From this derivation, it follows that $Q_{i,n}^k(0)$ in Eq.(\ref{eq:PCA_correction_term}) represents the linear response characteristics of the phase to deviations of $X_{i,n}^k$ from the state of $\phi_i=0$ on the limit-cycle solution.

%%%%%%%%%%%%%%%%%%%%%%%%%%%%%%%%%%%%%%%%%%%%%%%%%%%%%%%%%%%%%
%%%%%%%%%%%%%%%%%%%%%%%%%%%%%%%%%%%%%%%%%%%%%%%%%%%%%%%%%%%%%
%%%%%%%%%%%%%%%%%%%%%%%%%%%%%%%%%%%%%%%%%%%%%%%%%%%%%%%%%%%%%

%%%%%%%
% \bibliography{Reference}

%%%%%%%%%%%%%%%%%%%%%%%%%%%%%%%%%%%%%%%%%%%%%%%%%%%%%%%%%%%%%%%%%%%%%%%%%%%%%%%%%%%%%%%%%%%%%%%%%%%%%%%%%%%%%%%%%%%%%%%%%%%%%%%%%%%%%%%%%%%%%%%%%%%%%%%%%%%%%%%%%%%%%%%%%%%%%%%%%%%%%%%%%%%%%%%%%%%%%%%%%%%%

%apsrev4-2.bst 2019-01-14 (MD) hand-edited version of apsrev4-1.bst
%Control: key (0)
%Control: author (8) initials jnrlst
%Control: editor formatted (1) identically to author
%Control: production of article title (0) allowed
%Control: page (0) single
%Control: year (1) truncated
%Control: production of eprint (0) enabled
%

\end{document}